\documentstyle[psfig]{mn}

\def\etal{{\rm et al.\thinspace}}
\def\eg{{\it e.g.\ }}

\def\ie{{\it i.e.\ }}
\def\cf{{\it c.f.\ }}
\def\hi{H{\sc i}}
\def\hii{H{\sc ii}}

\def\halpha{{H$\alpha$}}
\def\nH{\hbox{$N_{\rm H}$}}
\def\Mdot{\hbox{$\dot M$}}

\def\h50{\hbox{$h_{50}$\,}}
\def\spose#1{\hbox to 0pt{#1\hss}}
\def\ltsimm{\mathrel{\spose{\lower 3pt\hbox{$\sim$}}
	\raise 2.0pt\hbox{$<$}}}
\def\ltsim{$\mathrel{\spose{\lower 3pt\hbox{$\sim$}}
	\raise 2.0pt\hbox{$<$}}$}
\def\gtsimm{\mathrel{\spose{\lower 3pt\hbox{$\sim$}}
	\raise 2.0pt\hbox{$>$}}}
\def\gtsim{$\mathrel{\spose{\lower 3pt\hbox{$\sim$}}
	\raise 2.0pt\hbox{$>$}}$}

\def\fract#1/#2{\leavevmode\kern.1em                   
   \raise.5ex\hbox{\the\scriptfont0 #1}\kern-.1em
   /\kern-.15em\lower.25ex\hbox{\the\scriptfont0 #2}}
\def\deg{\hbox{$^\circ$}}
\def\arcm{\hbox{$^\prime$}}
\def\arcs{\arcm\hskip -0.1em\arcm}
\def\counts{{\rm\thinspace counts}}
\def\fdg{\hbox{$.\!\!^\circ$}}

\def\fss{\hbox{$.\!\!^{\rm s}$}}
\def\counts{{\rm\thinspace counts}}

\def\farcs{\hbox{$.\!\!\arcs$}}
\def\Myr{{\rm\thinspace Myr}}
\def\cm{{\rm\thinspace cm}}
\def\erg{{\rm\thinspace erg}}

\def\K{{\rm\thinspace K}}
\def\keV{{\rm\thinspace keV}}
\def\km{{\rm\thinspace km}}
\def\nm{{\rm\thinspace nm}}
\def\kpc{{\rm\thinspace kpc}}
\def\Lsol{\hbox{$\thinspace L_{\odot}$}}
\def\Zsol{\hbox{$\thinspace Z_{\odot}$}}

\def\Mpc{{\rm\thinspace Mpc}}
\def\Msol{\hbox{$\thinspace M_{\odot}$}}
\def\pc{{\rm\thinspace pc}}
\def\s{{\rm\thinspace s}}

\def\yr{{\rm\thinspace yr}}

\def\ps{\hbox{\s$^{-1}\,$}}
\def\countsps{\hbox{$\counts\,\ps$}}
\def\pcc{\hbox{$\cm^{-3}\,$}}
\def\pyr{\hbox{$\yr^{-1}\,$}}

\def\pcm2{\hbox{$\cm^{-2}\,$}}
\def\ergpcm3ps{\hbox{$\erg\cm^{-3}\s^{-1}\,$}}
\def\ergps{\hbox{$\erg\s^{-1}\,$}}

\def\kmps{\hbox{$\km\s^{-1}\,$}}

\def\Lsolppc3{\hbox{$\Lsol\pc^{-3}\,$}}
\def\Msolppc3{\hbox{$\Msol\pc^{-3}\,$}}

\title[Multiple superbubbles in NGC 5253?]
{Multiple superbubbles in the starburst nucleus of NGC 5253?
-- Implications for mass loss from dwarf galaxies}
 
\author[David K.~Strickland and Ian R.~Stevens]
{David K.~Strickland\thanks{Present address: The Department of Physics \&
Astronomy, Johns Hopkins University, 3400 North Charles Street, Baltimore,
MD 21218, USA. E-mail: dks@eta.pha.jhu.edu} 
and Ian R.~Stevens\thanks{E-mail: irs@star.sr.bham.ac.uk} \\  
School of Physics and Astronomy, The University of
Birmingham, Edgbaston, Birmingham, B15 2TT, U.K.} 

\date{Accepted .....; Received .....; in original form .....}

\pagerange{\pageref{firstpage}--\pageref{lastpage}} 
\pubyear{1998}

\begin{document}

\maketitle
\label{firstpage}

\begin{abstract}
We obtained a long {\it ROSAT} HRI observation of the nearby dwarf starburst
and Wolf-Rayet galaxy NGC 5253. Our aim was to resolve the source of the
soft thermal X-ray emission seen by the {\it ROSAT} PSPC, proposed to be
a luminous superbubble by Martin \& Kennicutt (1995).
Instead of a single superbubble, we find a complex of at least five
sources of X-ray emission, associated with the massive clusters of young
stars at the centre of NGC 5253. The individual $0.1$ -- $2.4 \keV$ X-ray
luminosities of the components lie in the range $2$ -- 
$7 \times 10^{37} \ergps$. Three of the components are
statistically extended beyond the HRI PSF, the
largest having a FWHM of $8^{+10}_{-4}$ arcsec, equivalent to
$160^{+200}_{-80}\pc$.
We consider the origin of the observed X-ray emission, concentrating
on the sources expected to be associated with the starburst region:
superbubbles, supernovae, supernova remnants and massive X-ray binaries.
To assess the X-ray luminosity of a young superbubble blown by a single
massive cluster of stars we perform hydrodynamical simulations
with realistic time-varying mass and energy injection rates. 
The predicted soft X-ray luminosity for a superbubble blown by a
$M_{\star} = 10^{5} \Msol$ cluster during its Wolf-Rayet period
agrees very well with
the luminosities inferred for the extended components seen by the HRI.
We conclude that the extended X-ray components are most likely
young superbubbles blown by individual young clusters in the starburst region.
Although we do not rule out a contribution to the observed soft X-ray emission
by SNRs and massive X-ray binaries, we argue that the majority of the
emission comes from a few young superbubbles blown by the young stellar 
clusters in the starburst region. 
We discuss in  detail the implications of multiple superbubbles
on the efficiency
of mass and metal ejection from dwarf galaxies
by starburst driven galactic winds. We suggest the presence of multiple
stellar clusters in starbursting dwarf galaxies 
and the resulting multiple superbubbles will reduce the total ISM mass
ejected from from dwarf galaxies compared to the current models which
only consider the blowout of a single superbubble.
\end{abstract}


\section{Introduction}
\label{sec:introduction}

Starbursts are periods of enhanced star formation, generally in the 
central several hundred parsecs of a galaxy, that will exhaust
the available gas supply in significantly less than a Hubble time.
Starburst events in dwarf galaxies are particularly interesting
as there is the potential for major mass loss from the galaxy and the 
ejection of a significant 
fraction of the interstellar medium (ISM)
 due to superbubbles and galactic winds driven by
the pressure of thermalised supernova and stellar wind ejecta from the
massive stars created in the starburst. 

The basic model for a starburst driven wind (\cf Suchkov \etal 1994)
is conceptually simple:
The thermalised kinetic energy from massive stellar winds and SN ejecta 
in the starburst region create a hot, X-ray
emitting bubble in the ISM of host galaxy. This expands, sweeping
up and shock heating the ambient medium to form a superbubble. 
In a disk-like ISM this superbubble will break out of the disk after
a few million years to form a 
bipolar galactic wind, as seen around M82 (Watson, Stanger \& Griffiths
1984; Strickland, Ponman \& Stevens 1997). 
In dwarf galaxies with less disk-like ISMs, the
superbubble may sweep up and accelerate a significant fraction of the
ISM to escape velocity (De Young \& Heckman 1994) before blowing out into
the inter-galactic medium (IGM).

The loss of gas, and in particular
material enriched with heavy elements due to nucleosynthesis in 
massive stars, has a range of 
cosmological implications:
for the chemical evolution of dwarf galaxies
(Bradamante, Matteucci \& D'Ercole 1998 and references therein); 
their structure and evolution,
possibly turning dI galaxies into dE galaxies (see Marlowe \etal 1997 for
a detailed discussion) or disrupting them totally if they don't 
have massive dark haloes (Vader 1986; Dekel \& Silk 1986); and
for the chemical enrichment and heating of the Intra-Galactic medium 
(Spaans \& Norman 1997). 

The study of local starbursting dwarf galaxies has been invigorated
by the discovery of low mass star forming galaxies at intermediate redshifts.
Local starburst galaxies are readily accessible analogues of the star forming
galaxies at higher redshifts, allowing the processes of galaxy
evolution and formation to be investigated at a level of detail otherwise
unobtainable.
 
Understanding the effect of starbursts on low mass galaxies requires
quantitative measurements of the total energy and mass of gas 
(and newly synthesised heavy elements) incorporated into 
starburst-driven superbubbles
and galactic winds, and how much of this escapes the galaxy,
as a function of the properties
of the host galaxy and the star formation rate.

X-ray observations provide the only direct probe of the hot, metal enriched, 
gas that fills most of the volume of superbubbles and contains most of 
the energy. Most of the mechanical energy injected by the stellar winds and
supernovae into a superbubble is stored as thermal energy in the hot
bubble interior. This hot phase also contains the newly synthesised 
heavy elements. Dense gas swept up and shock-heated by superbubbles and winds
does emit optical emission lines, 
most notably H$\alpha$, but this is only an indirect probe of superbubbles and 
winds. The optical emission lines are predominantly excited by
photoionisation from the massive stars (Martin 1997), making it
much harder to study superbubbles with them. X-ray observations therefor
provide the best single way to study starburst-driven
superbubbles and galactic winds.

Characteristically, the X-ray emission from
superbubbles and galactic winds is soft, $kT \sim 0.5 \keV$. 
Although the specific energy of the supernova ejecta corresponds
to a temperature $kT \sim 10 \keV$ if efficiently thermalised,
hard X-ray emission from superbubbles and galactic winds will be much
weaker than the soft X-ray emission from gas at several million degrees Kelvin.
Thermal conduction and mass loading both introduce cooler and denser material
into the interior of superbubbles, which dominate the X-ray emission as
the luminosity is proportional to the square of the density 
(\cf Weaver \etal 1977; Hartquist \etal 1986). In galactic winds  
much of the soft X-ray emission is believed to come from shock-heated
clouds caught up in a faster moving, hotter 
but more tenuous wind (Suchkov \etal 1994).

Soft X-ray emission associated with the central regions and haloes 
of starburst galaxies
is therefore taken as evidence of starburst-driven bubbles or winds. 
Diffuse X-ray emission is seen extending beyond the optical disks of
some of the nearest starburst galaxies such as M82 (Strickland \etal 1997), 
NGC 1569 (Heckman \etal 1995) and NGC 4449 (Della Ceca, Griffiths \&
Heckman 1997), with properties
consistent with those expected of mature galactic winds.

The observational status of X-ray emission from younger starbursts,
where the superbubbles are still confined within the disk of the galaxy,
is more uncertain. 
{\it ROSAT} Position Sensitive Proportional Counter (PSPC) observations of
dwarf star forming galaxies such as NGC 5253 (Martin \& Kennicutt 1995),
I Zw 18 (Martin 1996) and NGC 4861 (Motch, Pakull \& Pietsch 1994) 
show unresolved soft X-ray sources coincident 
with their starburst regions. These are believed to be superbubbles
on the basis of their soft thermal spectra and association with the
starburst. Unfortunately, current X-ray instruments with the spectral-imaging
capabilities necessary to identify superbubbles have, at best, only moderate
spatial resolution. For example, the {\it ROSAT} PSPC's (the highest
spatial resolution spectral-imager to have been flown
at the time of writing)\footnote{The soon to be launched 
Advanced X-ray Astrophysics Facility
({\it AXAF}) will revolutionise this field, with a collecting area and 
spectral resolution far superior to {\it ROSAT}, and a spatial resolution
of $\sim0\farcs5$ (equivalent to $\sim 10 \pc$ at the distance of 
M82 or NGC 5253)!} 
spatial resolution of $\sim 30\arcs$
corresponds to a physical size of $\sim 600\pc$ 
at a distance of $4 \Mpc$, typical
of a {\em nearby} starburst galaxy such as NGC 5253 or M82.
This resolution is too low
to resolve any young superbubbles, and hence higher resolution X-ray 
observations are required to establish the exact origin of the emission.
Without high
resolution X-ray observations, the exact source of the soft X-ray emission is
unknown, and point source contamination of spectra also
becomes a potential problem.

This ambiguous observational status is unfortunate, as the superbubble
phase is important to study for several reasons:
\begin{enumerate}
\item The dynamics of superbubbles are less complicated than those of
mature galactic winds (\eg Suchkov \etal 1994), which are not as well understood
as those of superbubbles. As a result a quantitative understanding of 
superbubbles should be easier to achieve.
\item The superbubble phase is the best time to study the coupling between
the massive stars formed in the starburst and the ISM they affect through
superbubbles. Once the superbubble has blown out of the ISM into the
IGM the interaction between the hot gas from the starburst and the remaining
ISM is reduced, and as the wind expands into the halo the X-ray emission gets
fainter as the density drops. The parent stellar population of a 
young superbubble will also be easier to identify and study.
\item Studies of the superbubble phase are important for calibrating
and constraining
the theoretical models necessary to understand the later, more complex,
galactic wind phase and for obtaining quantitative estimates of 
the total mass and energy ejected from starburst galaxies.
X-ray observations cannot be used on their own, as they 
provide ambiguous results.  Standard single or two 
temperature spectral model fits to
X-ray data do not accurately represent the complex multi-temperature
gas distribution in superbubbles, as hence can not be used 
to accurately infer the true gas properties (see Strickland \& Stevens 1998)
on their own.
\end{enumerate}

We decided to observe the nearby dwarf starburst galaxy 
NGC 5253 with the High Resolution Imager (HRI) on board
{\it ROSAT}, with the aim of resolving and determining
the origin of the soft X-ray emission seen by the {\it ROSAT} PSPC,
believed to be a luminous superbubble (Martin \& Kennicutt 1995).

NGC 5253 is an ideal target for several reasons:
\begin{enumerate}
\item It is widely acknowledged to be a very young starburst. Rieke,
Lebofsky \& Walker (1988) place it as the youngest starburst in their 
chronological sequence of starbursts based upon its high Br$\gamma$
equivalent width. Its radio spectrum is almost entirely thermal, unlike many 
starburst galaxies, indicating very few classical supernova remnants 
(Beck \etal 1996).
Spectral features due to Wolf-Rayet (WR) stars have been identified in two
of the massive star clusters, constraining them to be only 
3 -- $5 \Myr$ old  (Schaerer \etal 1997).
\item Soft X-ray emission attributed to a luminous superbubble had 
already been discovered in the ROSAT PSPC observation of 
Martin \& Kennicutt (1995).
\item It is one of the nearest Wolf-Rayet galaxies (Conti 1991)
at a distance of $4.1 \Mpc$\footnote{The Wolf-Rayet galaxy NGC 1569 is
nearer than NGC 5253 at distance of only $2.2 \Mpc$,
 but is clearly an older and more evolved starburst
galaxy with a complex star formation history and a well
developed galactic wind (Vallenari \& Bomans 1996;
Gonz\'alez Delgado \etal 1997).} (Saha \etal 1995).
Wolf-Rayet galaxies are a class of young 
starburst galaxies, and are significantly X-ray overluminous
compared to a sample of ``normal'' galaxies and more mature starbursts
(Stevens \& Strickland 1998a, 1998b). Their {\it ROSAT} PSPC 
spectra are best characterised as emission from a soft thermal plasma.
One explanation for their excess X-ray luminosity is that it is due to young
superbubbles, which later blow out of the galaxy and fade.
\end{enumerate}

NGC 5253 therefore represents a chance to study at high resolution
(the $\sim 5 \arcs $ resolution of the HRI corresponds to a physical
size of $100\pc$ at the distance of NGC 5253)
the X-ray properties of a 
very young starbursting dwarf galaxy, with a wealth of supporting observations
at other wavelengths reported in the literature.

We shall briefly review previous X-ray observations of NGC 5253 in 
\S~\ref{sec:previous_xray}, before describing our {\it ROSAT} HRI analysis
in \S~\ref{sec:analysis} and our results in 
\S~\ref{sec:n5253_results}.
We detect at least five sources in the starburst region of NGC 5253, instead
of the single superbubble expected. The possible sources of the five X-ray
components are discussed in \S~\ref{sec:discussion}, concentrating on
superbubbles, supernovae, supernova remnants and massive X-ray binaries,
given the association between the X-ray emission and the young massive stars
in the starburst region. We conclude that multiple superbubbles blown by
the massive clusters of young, recently-formed stars probably dominate
the X-ray emission from NGC 5253. The implications of multiple
superbubbles for starburst-driven mass loss from NGC 5253 and other
dwarf galaxies are discussed
in \S~\ref{sec:implications}, before summarising our conclusions in 
\S~\ref{sec:n5253_conc}.

\section{Previous X-ray observations}
\label{sec:previous_xray}
NGC 5253 has previously been detected at X-ray wavelengths by both the
{\it Einstein} IPC and the {\it ROSAT} PSPC.

Using a $34$ ksec
observation with the  
{\it ROSAT} PSPC, Martin \& Kennicutt (1995) found a
single marginally extended ($\sim10\arcs$) source. At a distance
of $4.1 \Mpc$ the physical size of this possible extended source
is $\sim 200 \pc$.
The X-ray spectrum
was well fit with a soft, $kT \sim 0.4 \keV$,  absorbed thermal
plasma model, with a total $0.1$-$2.4\keV$  X-ray luminosity of
$L_{\rm X} \sim7 \times 10^{38} \ergps$.

A re-analysis of this data as
part of a {\it ROSAT} sample of WR galaxies by  Stevens \& Strickland
(1998) finds a slightly lower luminosity of $L_{\rm X} \sim 4\times
10^{38} \ergps$, due to using a lower value for the absorbing column
density.

 A previous $6$ ksec
observation with the lower resolution and sensitivity {\it Einstein} 
IPC detected a source coincident
with NGC 5253's position with a count rate that corresponds to a
$0.1$ -- $2.4 \keV$ X-ray luminosity of $L_{\rm X} \sim 5 \times 10^{38}
\ergps$, consistent with  {\it ROSAT} PSPC observations 
assuming the same spectral model.
(Fabbiano, Kim \& Trinchieri 1992).

Considering the various possible sources of the X-ray emission  Martin
\& Kennicutt (1995) provisionally attributed it to hot gas in a
superbubble, heated by supernovae and stellar winds from the  young
stars created in the starburst.

\section{HRI data reduction and analysis}
\label{sec:analysis}
NGC 5253 was observed by the {\it ROSAT} HRI (David \etal 1996) for
$71472 \s$ between the 25th to 28th, July 1996. The data
was reduced and analysed using the {\sc Asterix} X-ray analysis
package.  For a detailed description of the {\it ROSAT} HRI see David
\etal (1996) or Briel \etal (1994). After screening for periods of
poor pointing  accuracy (a negligible reduction)  and dead-time time
correction, the effective exposure time on axis is $71003 \s$.

The HRI has only a crude spectral sensitivity, although it is possible
to generate hardness ratios (Wilson \etal 1992). The pulse height
analyser (PHA) distribution of the HRI background, which is  mainly
particle events as opposed to the soft X-ray background, differs from
the PHA distribution of X-ray sources. To reduce the background we
used only data between PHA channels 3 and 8, inclusive. This maximises
the sensitivity for point sources and low surface brightness diffuse
features. Using only these channels reduces the background rate by
$\sim$ 25\% to $8.2 \times 10^{-7} \countsps {\rm arcsec}^{-2}$ from
the nominal rate of $1.1 \times 10^{-6} \countsps {\rm arcsec}^{-2}$.

\subsection{Sources and positional accuracy}
The absolute positions of {\it ROSAT} HRI sources are only accurate to
$\sim6\arcs$ (\cf Briel \etal 1994; David \etal 1996). 
This corresponds to an physical
uncertainty of $\sim120 \pc$ 
assuming the distance to NGC 5253 is $4.1\Mpc$. 
This uncertainty can be reduced if X-ray sources
within the field of view can be identified with objects at other
wavelengths, where usually the positions are known with better
accuracy (see, among many examples, Wang 1995). 

We used the maximum likelihood point source searching algorithm {\sc
Pss} to find sources more than $4\sigma$ above the background, over
the entire  HRI field of view. Several methods of estimating the
background were used, but all gave essentially the same resulting
source list,  apart from the most marginal detections. Excluding NGC
5253 itself, 31 sources were found.  We will discuss the sources
within the optical confines of NGC 5253 below.

We cross-correlated our HRI source list (excluding NGC 5253 itself)
with all sources within the same region of sky from SIMBAD,  the Nasa
Extragalactic Database (NED), the HST Guide Star Catalogue (GSC) and
those X-ray sources detected in the PSPC observation.  In addition we
cross-correlated the HRI source list with the positions of any object
seen, in a Digitised Sky Survey (DSS) image of the HRI field of view,  
that lay within $30\arcs$ of a HRI source.
 
No correlation with any source in the SIMBAD and NED databases was
found,  including the two historical SN Ia in NGC 5253. Five of the
sources were detected in the PSPC observation. The offsets and
bearings of the nearest optical source from the DSS image for each
X-ray source are random. 

For a maximum correlation distance of $20\arcs$, only one correlation
between our HRI source list and the GSC was found.
One of the brightest X-ray sources in the field of view, a $7\sigma$
detection (also seen in the PSPC observation) was offset from the
position of the Guide Star 072660005 ($\alpha_{2000.0} = 13^{\rm h} 39^{\rm m}
50\fss69$,  $\delta_{2000.0} = -31\deg 34\arcm 11\farcs6$) by  $\Delta \alpha
= +2\farcs7$, $\Delta \delta = +1\farcs8$.  

We have corrected the nominal HRI pointing solution to align the
X-ray source with the GSC star position. It is worth noting that with only
one cross-correlation we cannot assess if the nominal HRI pointing
solution is rotated with respect to the GSC.  We regard the absolute
X-ray positions quoted in this paper to be  more accurate than the
nominal HRI uncertainty of $\sim6\arcs$, although no more accurate
than $\sim3\arcs$.

\begin{figure}
\vspace{7.0cm} 
\includegraphics{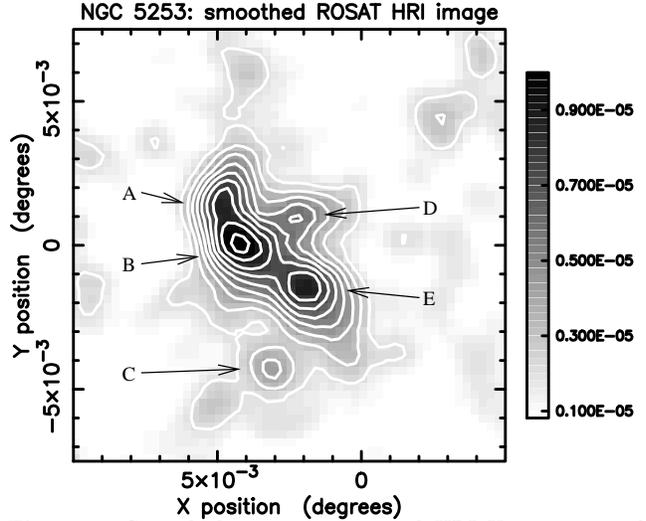}
\caption[Smoothed exposure corrected HRI X-ray image of NGC 5253]
 {Smoothed exposure corrected HRI X-ray image of NGC 5253. The
 data was binned in $1\arcs$ square pixels, and smoothed with a
 Gaussian mask of FWHM $4\farcs7$. Units are counts $\ps$
 arcsec$^{-2}$.  The letters refer to the X-ray components identified
 in the text and  in Table~\ref{tab:model_5ps}. The contour levels
 begin at a surface brightness of $2 \times 10^{-6} \counts \ps$ 
 arcsec$^{-2}$, and increase linearly in increments of $10^{-6} \counts \ps$
 arcsec$^{-2}$.}
\label{fig:n5253_ximage}
\end{figure}

\subsection{Image analysis}
\label{sec:images}
For the analysis of the soft X-ray emission from NGC 5253 itself  data
from the central $0\fdg 03 \times 0\fdg 03$ was binned into $1\arcs$
square  pixels, and corrected for vignetting, dead-time and exposure
time.

Visual inspection of the lightly smoothed image
(Fig.~\ref{fig:n5253_ximage}) reveals at least five, {\em apparently
point-like} sources of emission, over a $25\arcs \times 30\arcs$
($500\pc \times 600\pc$) region.

To quantify the X-ray emission we used a maximum likelihood image
fitting program to fit for the positions and intensities of five point
sources (modeled using the  functional form of the {\it ROSAT} HRI
point spread function [PSF]) and the intensity of a flat 
background to the
data. We used a maximum likelihood  method, rather than a traditional
$\chi^{2}$ technique,  given the low number of counts in each
component. 

To ease the fitting process, we supplied the fitting program a first
guess at the positions and  fluxes of the five components. The
background count rate at the position of NGC 5253 was extrapolated
from a polynomial fit to the
radial profile of the source-subtracted image of the entire HRI field
of view. The inner $0\fdg04$ was excluded to avoid biasing the
polynomial fit by any kiloparsec-scale  low surface brightness diffuse
emission that may be present near NGC 5253.  The estimated background
level was $2.95 \pm{0.10} \times 10^{-3}$  counts $\ps$
arcmin$^{-2}$. We shall return to the question of any larger scale
diffuse X-ray emission later, after first considering the five
observed components in more detail.

\begin{figure*}
\vspace{14cm}
\includegraphics{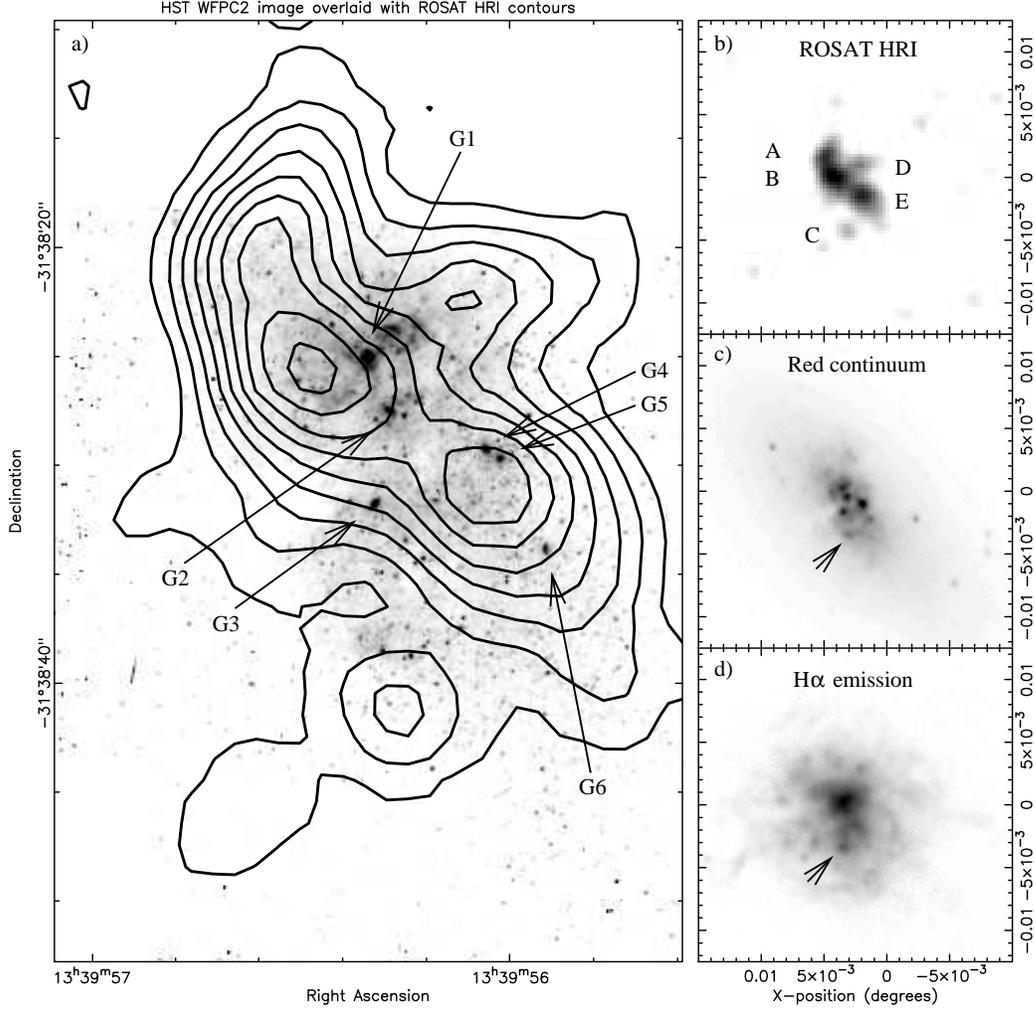}
\caption[HST WFPC2 image of the centre of NGC 5253 overlaid with a contour map  of the soft X-ray emission]
 {(a) A HST WFPC2 image of the centre of NGC 5253 (courtesy of
 V. Gorjian, 1996) overlaid with a lightly smoothed contour map of the soft
 X-ray emission seen by the {\it ROSAT} HRI (with the same contour levels as
 in Fig.~\ref{fig:n5253_ximage}).  The bright star clusters
 from Table~\ref{tab:cluster_mass} have been marked on the image.
 The intensity scale of the optical image has been adjusted to show
 only the young massive stars and clusters of the starburst region.
 The distribution of the X-ray emission with respect to the stars
 and ionised gas on a larger scale can be seen in (b) -- (d). These
 show on the same scale the X-ray emission (b), ground-based red
 continuum (c) and (d) \halpha~images (courtesy of C. Martin, see
 Martin \& Kennicutt 1995). The arrow in (c) \& (d) highlights the
 diffuse southern group of massive stars possibly associated with
 X-ray component C, discussed in \S~\ref{sec:images}.
 The absolute registration of the X-ray
 image with respect to the optical images is uncertain to $\sim3\arcs$,
 as discussed in \S~\ref{sec:analysis}.}
\label{fig:xhst}
\end{figure*}

To gain some insight into the origin of the X-ray emission, we have
overlaid a contoured HRI image on a HST WFPC2 image (see Fig.~\ref{fig:xhst}a,
courtesy of V. Gorjian, originally published in Gorjian 1996).
This shows that the soft X-ray emission 
comes from the starburst region near the
centre of NGC 5253, and the most intense emission is apparently
associated with the massive young stellar clusters. 

The brightest X-ray component B (identified in Fig.~\ref{fig:n5253_ximage}) is
only slightly offset to the east from the brightest cluster seen in the
optical. This cluster is bright enough to qualify as a ``Super Star Cluster''
(SSC, Meurer \etal 1995). SSCs are possibly the progenitors of globular
clusters, and are found in many starburst galaxies. Another bright cluster
of young stars is only $\sim 3\arcs$ to the SW of the SSC.

The second brightest X-ray component E is also almost on top of two
slightly less massive clusters. The properties of the optical clusters
are discussed in greater depth in \S~\ref{sec:superbubbles}.

The southern-most X-ray component C is also associated with a more diffuse
association of massive stars or lower mass clusters. The brightness of the
main clusters makes this difficult to show in Fig.~\ref{fig:xhst}a. This
southern grouping of massive stars is more visible in the ground based
optical images kindly supplied to us by C. Martin 
(Fig.~\ref{fig:xhst}c, first published in Martin \& Kennicutt 1995),
 or as a local peak in the \halpha~emission 
(Fig.~\ref{fig:xhst}d, Martin \& Kennicutt 1995, or
Fig.~1b of Calzetti \etal 1997). 

The northern-most components A \& D are not obviously associated with 
any bright clusters. Together with component B, components A \& D 
appear to make a broken ring-shaped
structure open to the north. We regard this as spurious, and choose to
consider them as three separate components. 

On the larger scale, the soft X-ray emission is clearly confined
well within the optical extent of the galaxy,
and is less extended than the $\sim 1 \kpc$ scale 
\halpha~filament complex (Fig.~\ref{fig:xhst}b -- d).

A comparison between this WFPC2 image and other optical, UV and radio
observations can be found in Gorjian (1996).
Readers may also find it useful to compare Fig.~\ref{fig:xhst} with
the optical and \halpha~images published in Calzetti \etal (1997).

The coordinate systems for the optical images in Fig.~\ref{fig:xhst}
have been adjusted so
that the main starburst clusters have the positions quoted in Calzetti
\etal's (1997) study of all the HST observations of NGC 5253.

\subsection{Source sizes}

As can be seen in Figs.~\ref{fig:n5253_ximage} \&~\ref{fig:xhst}a, 
some of the components seem physically larger and
more extended than others. In particular the peaks of components  C, E
\& B appear much broader than those of components A \& D.  In order to
quantify the level to which the HRI data constrain the sizes of the
individual components, we replaced the point-source  model for
individual X-ray components with a model consisting of a Gaussian
source blurred by the HRI PSF.

Can we realistically expect to constrain the size of the individual
components, given we only have between $\sim20$ and $\sim45$ counts
in each of the components, and any extension is similar to 
the size of the ROSAT HRI PSF 
(HWHM $\approx 2\farcs8$, David \etal 1996)?

  We performed a set of simulations to
assess whether it is possible  to resolve marginally extended sources
(\ie HWHM $= 1$ -- $5\arcs$), for source and background count rates
appropriate to our NGC 5253 observation. The Gaussian model, convolved
with the HRI PSF,  was used to produce simulated sources, of chosen
size (HWHM in the range $0\farcs5$ to $5\farcs0$)  and count rate, on
top of a flat background.  Adding Poisson noise, and then exposure
correcting, gave an artificial observation which was fitted with the
original model of a flat  background and Gaussian source blurred by
the HRI PSF. Several different Poisson realisations were made for each
size model.

\begin{figure*}
\vspace{9.0cm}
\includegraphics{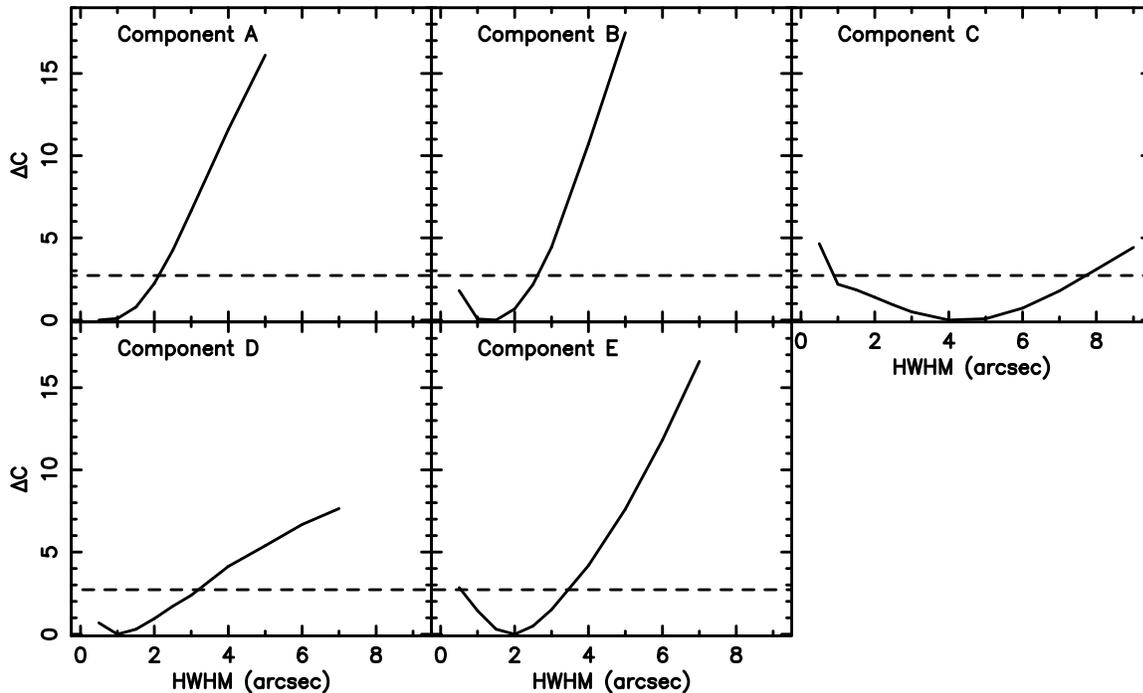}
\caption[Cash statistic as a function of source size]
 {The shape of the fit surface near the fit minimum for a
 Gaussian model (convolved with the HRI PSF) 
 fit to each X-ray component. The plots show the change in Cash
 C-statistic
 from the best-fit value, $\Delta C$ (equivalent to $\Delta \chi^{2}$), 
 as a function of the Gaussian
 HWHM. For each component, the models for the  other four components
 are fixed at their best-fit values (see the  text for
 details). Components A \& D are consistent with being unresolved, \ie
 point sources, while components B, C \& E are consistent with being
 marginally extended (HWHM in the range $1\farcs5$ to $4\arcs$)  with
 respect to the HRI PSF (which has a HWHM $\approx 2\farcs8$, 
 David \etal 1996). 
 The intersection of the dashed line shown at $\Delta C = 2.706$
 with the fit surface represents the
 90\% confidence interval in HWHM. }
\label{fig:extension}
\end{figure*}

For simulated sources having a similar total number of counts to the brighter 
X-ray components B \& E, it is possible to distinguish weakly resolved
(HWHM $\gtsimm 1\arcs$) from unresolved sources with a typical uncertainty
in HWHM of $\sim 1\arcs$. For fainter sources, similar to components C \& D, only
larger sources (HWHM $\gtsimm 3\arcs$) can be distinguished as being
inconsistent with the HRI PSF.

On this
basis, we expected components B, C \& E to be the best candidates for
showing any true source extension, given that components B \& E are
the brightest, and component C, although fainter, appears large and is
not in the wings of the brighter components as components A \& D
are. From the simulations we felt confident that our HRI data could be
used to distinguish  between truly unresolved point-sources and weakly
extended sources, for at least a few of the  observed components, and
place meaningful limits on their sizes.

Rather than attempt to directly fit five Gaussians and  a background
level to the data, we adopted the more ``hands-on'' method described
below, given the low count rates, complexity of the large fit
parameter space and potential for false minima.

We stepped individually through the components,  replacing the point
source model with  a Gaussian model at the point-source best-fit
position.  The other four components were fixed at their  best-fit
point source positions and fluxes. We then investigated the  variation
in best-fit Cash C-statistic (Cash 1979) with Gaussian size (as
parameterised by the HWHM, between $0\farcs5$ and $15\arcs$),  fitting
for position and both the Gaussian and the background-level
normalisation.

Only components B, C \& E showed any evidence for significant minima
at non-zero extension, with best-fit Cash statistic values that improved on the
best-fit five point source model. The best-fit total flux  for these
components also increased, indicating that the point
source model was missing flux.  However, replacing only one component
at a time with the Gaussian model, and leaving the remaining
components as point-sources is less than ideal, given that three
components showed signs of being extended. Some, although not all, of
the apparent extension could be due to the Gaussian model  picking up
flux from the other extended sources that the point source model
missed.  

We then replaced the best-fit point source models for components B, C
\& E with the best-fit Gaussian models, and then re-ran the procedure
detailed  above of stepping through all the components, replacing each
in turn  with a Gaussian model and investigating how the Cash
statistic varies  with Gaussian HWHM. As experiment showed that the
best-fit positions  were not altering, they were fixed at this point,
and not fitted for again.  Components A \& D remained statistically
unresolved, while components B \& C were reduced in size. Component E
remained almost the same size. The change in Cash statistic $\Delta
C$ (equivalent to $\Delta\chi^{2}$) from the minimum value found
as a function Gaussian HWHM for each component
is shown in Fig.~\ref{fig:extension}. 

After updating the current best-fit model (two point sources, three
Gaussians and the background normalisation), we finally fitted for the
normalisation of all the components, and the size of the three
Gaussian models, as we were confident that we had stepped as close to
the true fit minimum as possible. The final best-fit parameters and
90\% confidence regions for the five components are given in
Table~\ref{tab:model_5ps}. For components A \& D, we place upper
limits on their size (shown in Table~\ref{tab:model_5ps}) 
by estimating the Gaussian HWHM for a $\Delta C =
2.71$.

It is unlikely that the apparent extension of the X-ray components
is due to any systematic error in either our fitting process, or
a systematic blurring of the HRI PSF in the data (\eg the spatial
smearing discussed by Harris \etal 1998). If this were the case, we 
would expect to see all five components to show approximately
the same extension, which is clearly not the case.

\section{Results}
\label{sec:n5253_results}

\subsection{X-ray emission from within NGC 5253}

Assuming all the sources are spectrally similar, the best-fit spectral
model from the {\it ROSAT} PSPC observation can be used to convert HRI
count rates to X-ray luminosities.

As described in \S~\ref{sec:previous_xray}, Martin \& Kennicutt (1995) found
that the {\it ROSAT} PSPC spectrum was well described by a soft Raymond-Smith
hot plasma model (Raymond \& Smith 1977) 
of temperature $kT \approx 0.34\pm{0.10} \keV$, assuming
a total absorbing column of $\nH = 2.5 \times 10^{21} \pcm2$ (based on
examination of \hi~maps and the visual extinction towards to core on NGC 5253).
Stevens \& Strickland (1998a) got very similar results, $kT \approx 0.43 \keV$
and $\nH = 8^{+10}_{-4} \times 10^{20} \pcm2$ 
(fitting for the hydrogen column).

Assuming a Raymond-Smith hot plasma model for the X-ray emission, based
on the results quoted above, of
temperature $T=0.4\keV$ and metallicity $Z=0.25 Z_{\odot}$, and
correcting for a total absorbing column (\ie Galactic and that
intrinsic to NGC 5253)  of $\nH = 10^{21} \pcm2$, a $0.1$--$2.4$
luminosity of $L_{\rm X} = 1.15 \times 10^{41} \ergps$ corresponds to
one count per second in the {\it ROSAT} HRI.  For the Galactic column
of $\nH=4.0 \times 10^{20} \pcm2$ (Dickey \& Lockman 1990), 
one HRI count per second
corresponds to a luminosity of  $L_{\rm X} = 8.73 \times 10^{40}
\ergps$.  The inferred intrinsic X-ray luminosities are given in
Table~\ref{tab:model_5ps}, and range between $2$ -- $7 \times 10^{37} \ergps$
for the different components.
Note that the errors in $L_{\rm X}$ reflect
only the uncertainties in the count rates, and not those in the
spectral model itself. The count rate to luminosity conversions 
quoted above, and hence the inferred X-ray luminosities of the 
five X-ray components, are sensitive to
 the assumed spectrum and the
assumed column density.

We have assumed a typical absorption column of $\nH = 10^{21} \pcm2$ based
on the fitted column of Stevens \& Strickland (1998a), although the
absorption may be
a few times higher than that, as assumed by Martin \& Kennicutt (1995). 
The peak hydrogen column density in centre of NGC 5253, beam-averaged
over a region $\sim 1 \kpc$ in size, from 
the \hi~measurements of Kobulnicky \& Skillman (1995) is 
$\nH = 2.6 \times 10^{21} \pcm2$. If we assume the neutral hydrogen
is uniformly distributed in front of and behind the central starburst
region, the net absorbing column experienced by X-ray photons from the
core of NGC 5253 is $\nH \sim 1.7 \times 10^{21} \pcm2$. Note that the
HST-based study conducted by Calzetti \etal (1997) has shown that the 
optical extinction in the starburst region is very patchy, from regions
with almost no absorption to regions hidden behind 10 -- 30 magnitudes
of optical extinction (equivalent to hydrogen columns of 
$\nH \sim 2 \times 10^{22}$ -- $6 \times 10^{22} \pcm2$ based on the
conversion between hydrogen column and visual extinction published in
Gorenstein 1975).
However  
emission from extended sources of size a few arcseconds ($\gtsimm 100 \pc$),
such as components B, C \& E, is less affected by 
localised patches of dust and high extinction. In the absence
of more information regarding the origin and distribution of the
X-ray emission from NGC 5253 our assumed
hydrogen column  of $\nH = 10^{21} \pcm2$ is by no means unreasonable.

What limits can be placed on the X-ray luminosities of the individual
X-ray components seen in the HRI observation? Assuming only the
galactic hydrogen column, \ie no absorption
intrinsic to NGC 5253, gives a rough lower limit on the
$0.1$ -- $2.4 \keV$ X-ray luminosity (as given in Table~\ref{tab:model_5ps}).
This lower limit is only approximate, as it does not account
for the uncertainty in the true source spectrum for each source.
 
If $\nH \sim 2 \times 10^{21} \pcm2$ then the $0.1$ -- $2.4 \keV$ X-ray
luminosities of the individual X-ray components seen by the HRI are
a factor 1.44 times those given in Table~\ref{tab:model_5ps}. 
For a column of $\nH = 2 \times 10^{22} \pcm2$ the intrinsic X-ray
luminosities are nearly two orders of magnitude higher than those given
in Table~\ref{tab:model_5ps}, but note that such high columns are
inconsistent with the results of the spectral fits to the {\it ROSAT} 
PSPC spectrum, and the assumed soft thermal model over-estimates the
intrinsic X-ray luminosity if the true hydrogen column was as high
as high or higher than $2 \times 10^{22} \pcm2$. 

Is it safe to assume that all the sources have similar X-ray spectra?
The {\it ROSAT} HRI has a limited spectral capability (Wilson \etal
1992; David \etal 1996). Wilson \etal used the ratio of PHA channels
1-5 over 6-11 to define a softness ratio. We created smoothed maps in
channels 3-5  (soft) and 6-8 (hard), and created a hardness map,
(soft-hard)/(soft+hard).  With the low number of counts no evidence
for significant hardness variations between the X-ray emitting
components could be found.

\begin{figure}
\vspace{7.5cm}
\includegraphics{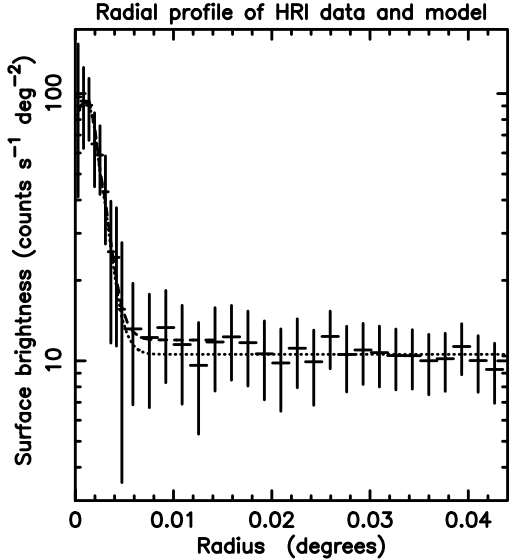}
\caption[A radial profile of the {\it ROSAT} HRI data]
 {A radial profile of the {\it ROSAT} HRI data, overlaid with
 the radial profile predicted by the best-fit model to the data (dashed)
 shown to a radius of $0\fdg015$, within which 
 the model was fitted. The dotted line shows the five component
 model with the expected background level, which clearly fits the data
 for $R \gtsimm 0\fdg02$. There is some evidence for excess flux above
 the expected background level for $R \ltsimm 0\fdg0.02$ ($\sim 1 \kpc$),
 similar to the maximum extent of the \halpha~emission.}
\label{fig:radial}
\end{figure}

\subsection{Extended diffuse emission}
Is there any evidence for  diffuse X-ray emission on larger scales, as
might  be expected given the kiloparsec-scale \halpha~filament complex
(Marlowe \etal 1995, Martin \& Kennicutt 1995)? Smoothed X-ray images
show no obvious diffuse emission above the noise
(Fig.~\ref{fig:n5253_ximage}), but the best-fit background level in the
central $0\fdg03 \times 0\fdg03$ is significantly higher than that
estimated using an source free annulus  and correcting for the HRI
vignetting function.

Using this background region, the estimated background level at the
centre of the field is  $2.95 \pm{0.10} \times 10^{-3}$ counts $\ps$
arcmin$^{-2}$.

From the image fitting, there is a statistically significant  excess
(at the $3\sigma$ level) of $12.3\pm{0.9} \times 10^{-4}$  counts
$\ps$ ($87\pm{6}$ counts) not accounted for by  expected background
and the five point sources themselves.  This corresponds to an
intrinsic X-ray luminosity of  $1.1\pm{0.1} \times 10^{38} \ergps$,
assuming $kT=0.4 \keV$, $Z=0.25 \Zsol$  and $\nH=4 \times 10^{20}
\pcm2$.

This is too large an excess to be explained by any inadequacies of our
PSF model for the HRI. It is difficult to assess whether the excess 
flux is associated with a
genuine  low surface brightness diffuse component, or is flux missed
from the central point-like components by using only a  five component
fit to the data. A radial profile centred on the saddle point between 
components B \& E does seem to suggest excess diffuse emission extending
out to a radius of $\sim 0\fdg015$ ($\sim 1 \kpc$).
A combination of future {\it
AXAF} and {\it XMM}  observations will tell us more.

\begin{table*}
\begin{minipage}{140mm}
\caption[Best-fit source parameters for the NGC 5253 HRI data]
  {Best-fit parameters for the NGC 5253 HRI data.  Errors quoted
  are 90\% confidence in one parameter of interest for count rates and
  size.}
\label{tab:model_5ps}
\begin{center}
\begin{tabular}{lcccccc}
\hline
Comp.  & $\alpha$ (2000.0)  & $\delta$ (2000.0)  & Count rate &
        $L_{\rm X}^{a}$ & $L_{\rm X}^{b}$  & size (HWHM) \\ &    &
        & ($10^{-4}$ counts $\ps$)   & \multicolumn{2}{c}{($10^{37}
        \ergps$)}  & (arcsec)$^{c,d}$ \\ \hline A         & $13^{\rm
        h} 39^{\rm m} 56\fss52$ & $-31\deg 38\arcm 17\farcs5$ &
        $4.1^{+1.5}_{-1.5}$ & $3.6^{+1.3}_{-1.3}$  &
        $4.7^{+1.7}_{-1.7}$ & $<2.0$ \\  B         & $13^{\rm h}
        39^{\rm m} 56\fss35$ & $-31\deg 38\arcm 24\farcs7$ &
        $6.3^{+2.1}_{-1.9}$ & $5.5^{+1.8}_{-1.7}$  &
        $7.3^{+2.4}_{-2.2}$ & $1.4^{+1.3}_{-1.4}$ \\  C         &
        $13^{\rm h} 39^{\rm m} 56\fss14$ & $-31\deg 38\arcm
        40\farcs2$ & $3.4^{+2.6}_{-1.8}$ & $2.9^{+2.3}_{-1.6}$  &
        $3.9^{+3.0}_{-2.1}$ & $4.2^{+4.9}_{-2.3}$ \\  D         &
        $13^{\rm h} 39^{\rm m} 55\fss75$ & $-31\deg 38\arcm
        20\farcs8$ & $2.4^{+1.3}_{-1.2}$ & $2.1^{+1.1}_{-1.1}$  &
        $2.7^{+1.5}_{-1.4}$ & $<3.0$ \\  E         & $13^{\rm h}
        39^{\rm m} 55\fss68$ & $-31\deg 38\arcm 29\farcs8$ &
        $6.3^{+2.0}_{-2.0}$ & $5.5^{+1.8}_{-1.7}$  &
        $7.3^{+2.3}_{-2.2}$ & $1.9^{+1.1}_{-1.3}$ \\  \hline
\end{tabular}
\end{center}
(a) Intrinsic $0.1$--$2.4\keV$ luminosity assuming a Raymond-Smith
hot plasma model with $kT=0.4 \keV$, metallicity  $Z=0.25 Z_{\odot}$,
corrected for a Galactic absorbing column of  $\nH=4\times10^{20}
\pcm2$. \\ (b) As (a) except for a total $\nH=10^{21} \pcm2$. \\ (c)
At a distance of $4.1\Mpc$, 1 arcsecond corresponds to $20\pc$. \\ (d)
Positions and fluxes for components A \& D  are from a point source
fit, with upper limits on size based on a Gaussian model
fit. Otherwise best fit parameters are from fits to the data of a
Gaussian source  blurred by the HRI point spread function.
\end{minipage}
\end{table*}

\subsection{Summary of observational results}
\label{sec:summary_results}
In summary the HRI observations show several, at least five, separate
sources of soft, 0.1-2.4 $\keV$, X-ray emission, associated with the
recent star formation at the centre of NGC 5253. The integrated X-ray
spectrum observed by the {\it ROSAT} PSPC is best characterised as
a soft thermal plasma, absorbed by a total column in excess of the
galactic hydrogen column, and with possible sub-solar metal
abundance. Within {\it ROSAT}'s pointing accuracy the brightest X-ray
components are associated with the four brightest young clusters of
massive stars, and the southern-most X-ray component is close to a
more diffuse association of massive stars. Based on the PSPC spectrum
the luminosities of the five components range between  $L_{\rm X}
\approx 2$ -- $7 \times 10^{37} \ergps$. Three of the components are
statistically extended beyond the HRI PSF at 90\% confidence,  the
largest having a FWHM of $8^{+10}_{-4}$ arcsec, equivalent to
$160^{+200}_{-80}\pc$.   The upper limits on the sizes of the other
two X-ray components are consistent with the sizes observed for the
two brightest components that are extended, so we cannot
rule out the possibility that they are  marginally extended as well.

\section{Discussion}
\label{sec:discussion}
Martin \& Kennicutt (1995), in their study of NGC 5253 using the {\it
ROSAT} PSPC, argued that the most likely source of the X-ray emission
was from a luminous superbubble evolving in a cloudy medium. A {\it
ROSAT} PSPC survey of 14 Wolf-Rayet galaxies (including NGC 5253) by
Stevens  \& Strickland (1998a; 1998b), also concluded that the source of the
X-ray  emission in these objects was most probably young superbubbles.

Given that with higher resolution the X-ray emission comes from
several components it is worth re-assessing the possible sources of
the observed emission.

\subsection{X-ray emission from the massive stars}
Individual O stars typically emit a fraction 
$L_{\rm X}/L_{\rm Bol} \sim 3 \times 10^{-7}$ of their
total bolometric luminosity in soft X-rays, although with
a scatter of about an order of magnitude around this mean value (Sciortino
\etal 1990). 

Using published data from the literature we can 
estimate the X-ray luminosities of the massive clusters
of young stars in NGC 5253 due to soft X-ray emission from the
massive stars alone. Table~\ref{tab:cluster_mass}
collates the observed properties of the six brightest clusters
in NGC 5253 (the brightest an example of a Super Star Cluster) from
the work of Meurer \etal (1995), Gorjian (1996), Calzetti \etal (1997) and
Schaerer \etal (1997).

From the estimated bolometric luminosities $L_{\rm bol} \sim 10^{40}$
-- $3 \times 10^{42} \ergps$ it is clear that individual massive stars
cannot be the source of the soft X-ray emission seen in NGC 5253.

\begin{table*}
\begin{minipage}{140mm}
\caption[Properties of the six brightest SSCs in NGC 5253]
  {Properties of the six brightest SSCs in NGC 5253, taken from
  Meurer \etal (1995, M95), Gorjian (1996, G96), Calzetti \etal  (1997,
  C97) and Schaerer \etal (1997, S97).  Masses  quoted are the total
  initial mass between the two mass limits assuming a Salpeter IMF. The names
  and numbers used here and in Fig.~\ref{fig:xhst} 
  have been prefixed by the first letter of the first
  author's name to reduce confusion between their different naming
  conventions. We have used the Leitherer \& Heckman (1995, LH95) instantaneous
  starburst models (for a Salpeter IMF and a metal abundance of 0.25 Solar)
  to estimate the bolometric luminosities and some of the masses.}
\label{tab:cluster_mass}
\begin{center}
\begin{tabular}{lcccccccc}
\hline
Parameter       & \multicolumn{6}{c}{Cluster name} & Ref. & Notes \\ & G1
                & G2 & G3 & G4 & G5 & G6   & G96 & \\ & SA & SB &    &
                &    &      & S97 & \\ & C5 & C4 & C1 & C2 & C3 & C6
                & C97 & \\ \hline Age (Myr)       & 2.8 & 4.4 &  &
                &    &      & S97 & \\ $\log N_{\rm Lyc} (\ps)$  &
                $52.0$ & $51.4$ &&&&& S97 & a \\ $N_{\rm O}$     &
                $\sim1700$         & $\sim840$ &&&&& S97 & b \\
                $N_{\rm WR}$    & $\sim35$           & $\sim52$ &&&&&
                S97 & c \\ $\log M_{\star}$ (0.1--100$\Msol$)  & $6.0$ & $5.7$
                &&&&& & d,e \\ & & &&&&& & \\ $R$ ($\pc$)     & 3.3 &
                3.0 & 3.1 & 2.2 & 1.5 & 1.7 & G96 & \\ & & &&&&& & \\
                Age (Myr)       & 2.2--2.8 & 4.0--4.4  & 8--12 &
                50--60 & 30--50 & 10--17 & C97 & f,h \\ 
$\log M_{\star}$                
        (0.1--100$\Msol$) & 5.5--6.0 & 4.0--4.6 & 4.8--5.6 &
                4.8--5.6 & 4.8--5.6 & 4.8--5.6 & C97 & d,i \\
$\log L_{\rm bol}$ ($\ergps$)
        & 42.5 & 41.5 & 41.0 & 40.0 & 40.2 & 40.8 & & j \\  
& &
                &&&&& & \\ $\log M_{\star}$ (0.1--100$\Msol$) & $\geq 3.8$ &
                $\geq 4.4$ & $\geq 4.3$ & $\geq 3.8$ & $\geq 3.9$ &
                $\geq 4.0$ & M95 & d,k \\ \hline
\end{tabular}
\end{center}
(a) Ionising Lyman continuum flux estimated from $H\beta$.
Extinction dependent. \\ (b) Total number of O stars. \\ (c) Total
number of Wolf-Rayet (WR) stars. \\ (d) Total mass in stars with
masses between 0.1 and $100\Msol$ assuming a Salpeter IMF.\\ (e)
Estimated here using inferred ages and number  of O stars with the
LH95 models.\\  (f) For clusters G1 \& G2 the estimated age is based
on  $H\alpha$ \& $H\beta$ equivalent widths. \\ (h) Age estimates
based on cluster colours. \\ (i) Based on de-reddened absolute
magnitude. \\ 
(j) Calculated from the LH95 models, assuming an initial 
mass of $10^{6}\Msol$ for G1 and $10^{5}\Msol$
for G2 -- G6, and the youngest age estimate for each cluster. \\
(k) Assuming clusters are at peak UV brightness, and
uniform reddening \\
\end{minipage}
\end{table*}

\subsection{Massive X-ray binaries}
Massive X-ray binaries (MXRBs) generally have hard X-ray spectra.
That NGC 5253's X-ray spectrum is well fit by a soft thermal plasma model
indicates that the majority of the X-ray emission
is not produced by MXRBs. 

MXRBs with a black hole (BH) rather than  a neutron star are a
possible source of the X-ray emission from the non-extended
components,  as they have  a soft spectral component in addition to a
hard power law tail,  and can have luminosities up to  $L_{\rm X} \sim
10^{38} \ergps$. The two strongest X-ray point sources  within 30
Doradus may be Wolf-Rayet + BH  binaries (Wang 1995).

The extension of X-ray components B, C \& E and their dominance of the
total count rate argue against all their emission being due to MXRBs
and/or BH-binaries. MXRBs and BH-binaries could well provide some
fraction of the total X-ray emission from NGC 5253. Higher resolution
observations with the $0\farcs5$ resolution of {\it AXAF} will be
necessary to identify what fraction of the soft X-ray emission comes
from point sources.

\subsection{Low-mass X-ray binaries}
Low mass X-ray binaries (LMXBs), although having softer X-ray spectra
than MXRBs, are still generally harder than the $kT\sim0.4\keV$
emission seen by the {\it ROSAT} PSPC. In addition,  the association
between the HRI X-ray sources and the clusters of young massive stars
argues against LMXBs being the sources of the observed emission.

\subsection{Supernovae and Supernova Remnants}
\label{sec:supernovae}
A natural consequence of the starburst phenomenon is a high supernova
(SN) rate for tens of millions of years within a relatively small
volume of the host galaxy.

SN occurring within cavities or bubbles blown by preceding SN or
stellar winds are best considered within the framework of the
superbubble model, discussed at length in \S~\ref{sec:superbubbles}. 
Individual SN and SN
remnants (SNR) can be luminous X-ray  sources in their own right,
which we shall consider here.

We can estimate the SN rate in NGC 5253 from an assumed star formation
history.  Meurer \etal (1995) estimate a total initial mass of $4.4
\times 10^{6} \Msol$  (for a Salpeter IMF extending between $0.1$ --
$100 \Msol$) based on the average mass-to-light ratio for a constant star
formation rate with an age between $1$ -- $100 \Myr$. For ages between $10$
-- $50 \Myr$ the resulting SN rate is $\sim 10 ^{-3} \pyr$ and is not
very sensitive to the star  formation history. 

No type II SN have been observed in NGC 5253, although two SN Ia have
occurred in the outer parts of the galaxy (Saha \etal 1995). The radio spectrum
of  NGC 5253 places upper limits on the total number of classical SN or SNR, 
as it is unusual for a starburst galaxy in 
being almost entirely thermal (90\% -- 95\%)
in origin, suggesting very few if any isolated SN \& SNRs (Beck \etal 1996)
These measurements would have detected the presence of 10 or more
SNRs, or a single radio SN.  The arcsecond resolution radio
observations show only one compact (but not point-like) radio
source,  associated with the very young and bright cluster G1
(adopting the naming convention from Table~\ref{tab:cluster_mass}).

Young (ages measured in decades) type II SN have a wide range of X-ray
luminosities, from  $L_{\rm X} \sim 2 \times 10^{35} \ergps$ for SN1987A
in the {\it ROSAT} band, to several $\times 10^{40} \ergps$ for
SN1986J and SN1993J  (Schlegel 1995 and references therein). With no
SN IIs observed within the last century it seems unlikely that any of
the observed X-ray emission is due to recent SN.

In general SNRs are faint X-ray sources, with $0.1$ -- $2.4 \keV$
luminosities and radii typically in the range  $10^{34}$ -- $10^{36}
\ergps$ and $10$ -- $30 \pc$ for remnants with soft thermal spectra
(\eg Smith \etal 1994; Rho 1996).  If any of the components seen by
the HRI are supernova remnants then they would be, by definition, the
most luminous SNR in NGC 5253, and hence not typical. A more
informative comparison is with the most luminous thermal X-ray
emitting remnants in the LMC: N123D, N63A \& N49 (Hughes, Hayashi \&
Koyama 1998).  These have $0.5$ -- $4.0 \keV$ luminosities of 3, 2 and
$0.6 \times 10^{37} \ergps$ respectively (luminosities for the
observed HRI components in the equivalent band are $\sim$ 60\% of
their $0.1$ -- $2.4 \keV$ luminosities). These middle aged 
(4 -- $6 \times 10^{3} \yr$) X-ray luminous remnants
have radii on the low end of the range quoted above, \eg $R \approx 8
\pc$ for N49 to $R \approx 12 \pc$ for N132D. The inferred ambient ISM
densities for these three SNR, between $2$ -- $4~{\rm H} \pcc$, are
the highest of Hughes \etal's (1998) sample of the seven brightest
thermal X-ray emitting SNRs in the LMC.

With an estimated SN rate of $10^{-3} \pyr$ for the starburst
in NGC 5253, one or more of the
observed components could be luminous, middle aged ($\sim 5000 \yr$)
SNRs similar to those seen in the LMC if the ambient density in their
vicinity was reasonably high. 
A wide range of observations  (\halpha:
Martin 1997, CO: Turner, Beck \& Hurt 1997,  strong and patchy
absorption and dust lanes: Calzetti \etal 1997)  demonstrate the
presence of dense gas in the starburst region.  The main problem with
individual luminous SNRs being the  source of the components seen by
the HRI is trying to explain the extended components. The luminous
SNRs in the LMC are significantly smaller ($R_{\rm SNR} \sim 10 \pc$)
than the estimated sizes  for the extended components (HWHM in the
range $\sim 30$ to $\sim80 \pc$).   A lower ambient density can give
larger remnants, but also reduces their X-ray luminosity. One solution
is to argue that the measured extension is a consequence of multiple
sources within a similarly sized region, in which their sizes and
individual luminosities can be smaller.  This model, while a
possibility, is beginning to require a number of reasonably luminous
thermal X-ray emitting SNRs that may be unrealistic given the observed
limits of $\ltsimm 10$ SNRs based on the radio spectrum and the SN
rate estimated above, if we are to explain all the apparently extended
X-ray components as supernova remnants.

\subsection{Superbubbles}
\label{sec:superbubbles}
Superbubbles are pressure driven bubbles in the ISM, inflated by the
combined mechanical energy input from stellar winds and SNe from the
population massive stars within them.  We refer the reader to the
extensive literature on the subject for a more detailed background
(see McCray \& Kafatos 1987;  Mac Low \& McCray 1988; Tenorio-Tagle \&
Bodenheimer 1988;  Mac Low, McCray \& Norman 1989; Chu \& Mac Low
1990; Wang \& Helfand 1991; Shull \& Saken 1995 among others). 

\begin{figure}
\vspace{8cm}
\includegraphics{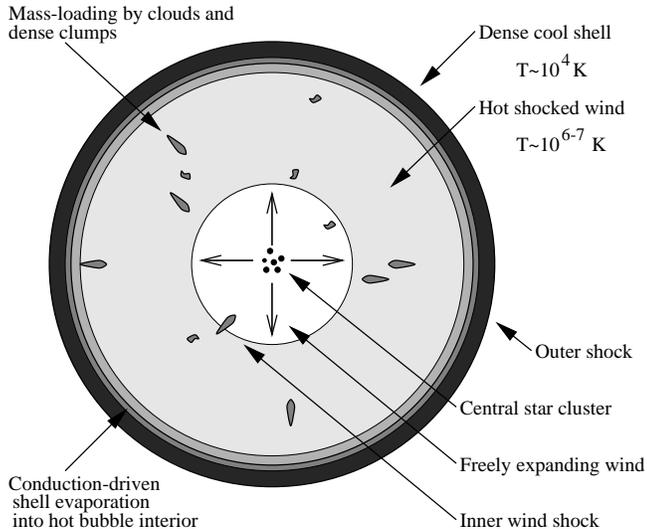}
\caption[A schematic picture of a superbubble]
 {A schematic picture of a superbubble.}
\label{fig:bubble_struct}
\end{figure}

Fig.~\ref{fig:bubble_struct} shows an idealised model for the
structure of a superbubble. A population of massive stars inject
kinetic energy and mass from stellar winds and SNe in the central star
cluster. It is assumed that shocks efficiently thermalise this energy,
creating a very hot, high pressure reservoir of gas ($T \sim 10^{8}
\K$ for typical values of  $L_{\rm w}$ and $\Mdot_{\rm w}$, assuming no
significant initial radiative energy losses and no additional mass loading from
clouds). This drives an outflow that becomes supersonic outside the
mass and energy injection region.  The structure of the central region
and the free wind are given analytically by Chevalier \& Clegg
(1985). The freely expanding wind is surrounded  by a region of hot,
shocked wind, which occupies the majority  of the volume of the
superbubble.  The bubble interior is the source of the X-ray emission
from the  superbubble. A contact discontinuity separates the hot bubble
interior from a shell of swept-up and shocked ISM. The cooling time of
this shell is typically short compared to the expansion  time scale of
the bubble, leading to a cool, $T \sim 10^{4} \K$, dense shell. The
outer edge of the shell is a shock wave propagating into the
undisturbed ISM surrounding the bubble. 

In this simple description, all the mass in the hot bubble interior is
supplied from the mass loss from the central star cluster, and all the
mass of the swept up ISM is confined to the cool shell. In practice,
additional mass enters the hot shocked wind, enhancing its density and
cooling the interior. In the absence of magnetic  fields, thermal
electrons from the hot interior cross the contact  discontinuity and
heat the edge of the cold shell. This thermal conduction leads to
the evaporation of a
small fraction of the cold shell mass into the bubble's  interior,
which can significantly enhance the density within the bubble
interior. Another source of mass for the bubble interior is
hydrodynamical ablation and destruction (``mass-loading'') of dense
interstellar  clumps and clouds that have been overrun by the bubble
(see Hartquist \etal 1986; Klein, McKee \& Colella 1994;  Arthur, Dyson \&
Hartquist 1993 and references therein), or their evaporation
due to thermal conduction (Cowie \& McKee 1977).
Sources of additional mass in the hot
bubble are of particular importance when considering the X-ray
emission from superbubbles, as X-ray emission is a two body process,
and the luminosity depends on the density squared.

Superbubbles are physically very similar to wind-blown bubbles around
individual massive stars, apart from the slightly different  mechanism
of mass and energy injection. The models of  Castor \etal (1975)
and Weaver \etal (1977; henceforth called the  Weaver model) for
such bubbles around single massive stars form the basis for the
analytical  treatment of superbubbles, and are commonly applied
directly to superbubbles without any modification.

The Weaver model is a one-dimensional similarity solution for the
structure and evolution of a wind-blown bubble of {\em constant} mass
and  energy injection, expanding into a homogeneous ISM of uniform
density. The solution applies for times when radiative cooling of the
tenuous,  hot interior of the bubble is negligible, but once the swept
up ISM has cooled and collapsed to form a thin, cool ($T\sim10^{4}
\K$) shell.  The Weaver model incorporates thermal conduction between
the hot  ($T \sim 10^{7} \K$) bubble interior and the cold shell,
leading to  evaporation of  material off the shell, cooling and
enhancing the density  within the bubble interior. For convenience we
reproduce two relationships for the radius of the contact
discontinuity (Castor \etal 1975),
\begin{equation} 
R_{\rm c} = 0.76 \, (L_{\rm w}/\rho_{0})^{1/5} \, t^{3/5},
\label{equ:n5253_radius}
\end{equation}
and the X-ray luminosity, incorporating the effects of conduction
(Chu \& Mac Low 1990),
\begin{equation}
L_{\rm X} \propto \Lambda_{\rm X}(T,Z) \, \rho_{0}^{17/35} \, L_{\rm
w}^{33/35} \, t^{19/35}.
\label{equ:n5253_lx}
\end{equation}
$L_{\rm w}$ is the mechanical energy injection rate
due to SNe and stellar winds, $\rho_{0}$ the ambient
density, $t$ the age of the bubble and $\Lambda_{\rm X}(T,Z)$ the
X-ray emissivity, which is a function of temperature $T$ and
metallicity $Z$.

\begin{figure*}
\vspace{13.5cm}
\includegraphics{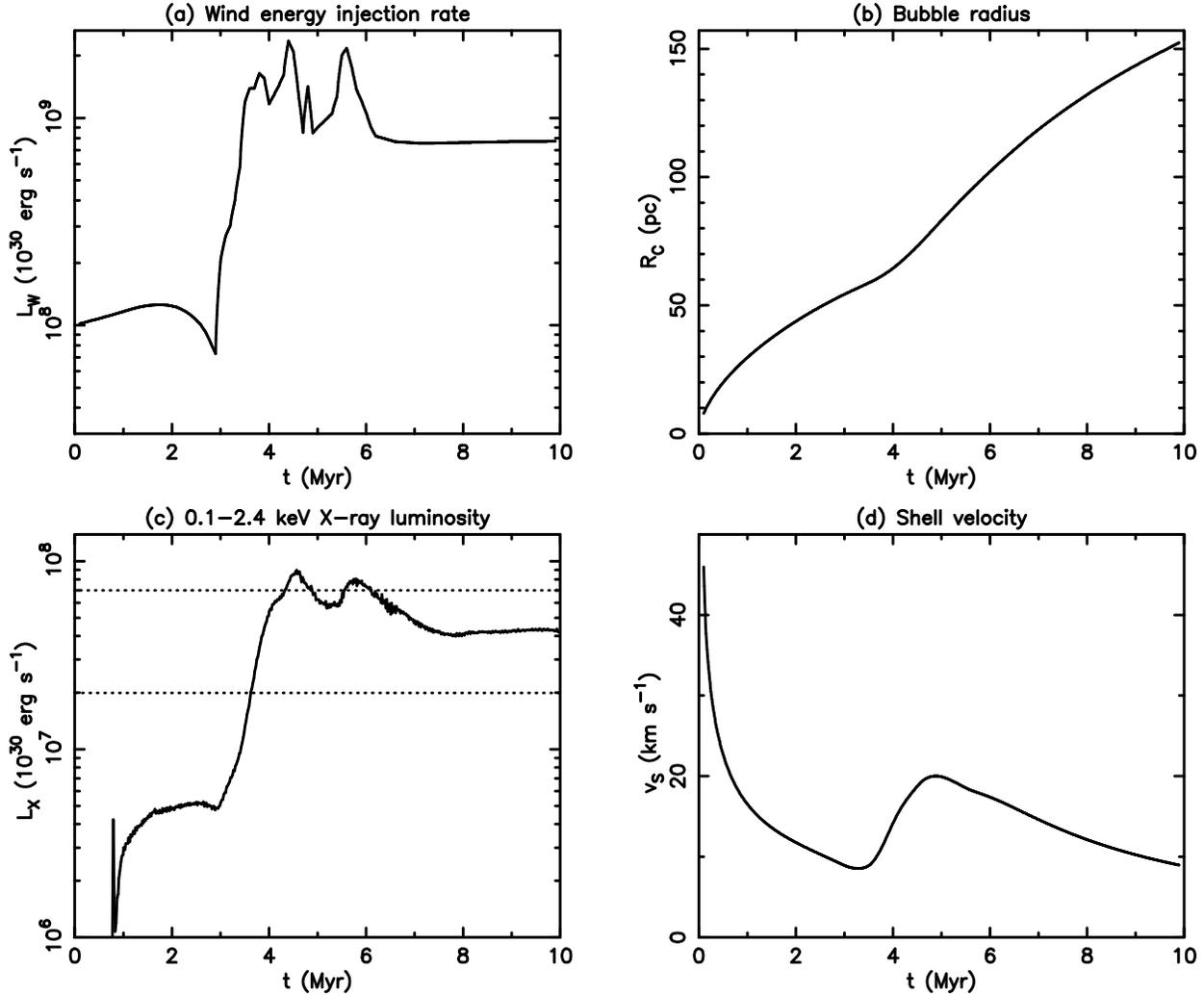}
\caption[Properties of a superbubble blown by a young stellar
 cluster of total mass
 $10^{5} \Msol$]
 {Properties of a superbubble blown by a cluster of total mass
 between $0.1$ --  $100\Msol$ of $10^{5} \Msol$.  (a) Mechanical energy
 injection rate $L_{\rm W}$ as a function of time, from Leitherer \&
 Heckman's (1995) starburst models, assuming an  instantaneous burst, a
 Salpeter IMF and a metal abundance of 0.25 Solar.  (b) Radius of the
 contact discontinuity separating the hot bubble from the cool shell of
 swept up ISM, as a function of time. This is the radius to which X-ray
 emitting gas extends.  (c) X-ray luminosity as a function of time in
 the  {\it ROSAT} $0.1$ -- $2.4 \keV$ band from the hydrodynamic
 simulation described in \S~\ref{sec:bubsim}.  
 Note the similarity in shape to the mechanical energy injection
 rate $L_{\rm W}$. The dashed lines show the range of X-ray luminosities
 inferred for the individual X-ray components given in
 Table~\ref{tab:model_5ps}.
 (d) Shell velocity $v_{\rm s}$. 
 The shell is Rayleigh-Taylor unstable
 between $t \approx 3.5$ and $5 \Myr$ as it is accelerated by the
 sudden increase in pressure in the bubble interior from the first SNe.}
\label{fig:lh95}
\end{figure*}

The Weaver model has widely been used in interpreting the X-ray
emission from wind-blown bubbles and superbubbles in the Galaxy (\eg 
Wrigge, Wendker \& Wisotzki 1994),  
and extragalactic bubbles (Martin \& Kennicutt 1995;  Heckman \etal 1995;
Stevens \& Strickland 1998a), in particular in the LMC
(Chu \& Mac Low 1990; Wang \& Helfand 1991; Oey \& Massey 1995). 

The main limitation of the Weaver model is its assumption of a
constant energy injection rate, and a single phase,  uniform density
ambient medium, neither of which apply in reality.  As can be seen
from Fig.~\ref{fig:lh95}a, the mechanical energy injection rate from a
young, co-eval population of massive stars is a strongly varying
function of time. This, as we shall show below, has a strong effect on
the soft X-ray luminosity as a function of time.


A variety of techniques have been developed to explore more complex
situations than the standard Weaver model allows:
\begin{enumerate}
\item Analytical models: 
Several workers (\eg Koo \& McKee 1992; Ostriker \& McKee 1988;  Zhekov \&
Perinotto 1996) have extended Weaver \etal's (1977) similarity
solution from uniform density and energy injection rates to power law
density distributions and energy injection rates. These solutions can even
incorporate conduction (Zhekov \& Perinotto 1996), assuming that the
time-scale for the bubble to come into conductive equilibrium is short
compared to the expansion time-scale.

For our purposes these solutions are less than ideal, as the
time-dependent mass and energy injection from a young cluster in a
starburst region is not well approximated by a power law in time
(Fig.~\ref{fig:lh95}a).  Analytical models also fail to
incorporate the hydrodynamical  instabilities that can be crucial in
shaping the dynamics and properties of these bubbles (see the
hydrodynamic simulations of wind-blown bubbles by Garc\'{\i}a-Segura,
Mac Low \& Langer 1996).

\item Semi-analytic numerical models:
If only the bubble's radius and velocity are required, an elegant
solution not requiring detailed numerical hydrodynamics, and one that can be
used for a wider range of density distributions and energy injection rates, is
to use the Kompaneets thin-shell approximation (see
Mac Low \& McCray 1988; Shull \& Saken 1995; Oey \&
Massey 1995). This does not provide any information regarding the
bubble interior other than its pressure, and hence can not be used
to explore its X-ray properties. 

An approximation employed by
Mac Low \& McCray (1988) and Chu \& Mac Low (1990) is to assume that
as each instant the radial density and temperature within the bubble obey
the steady-state conduction dominated similarity solution of Weaver \etal 
(1977):
\begin{equation}
T(r) = T_{\rm cen} ( 1 - r/R_{\rm c})^{2/5},
\end{equation}
\begin{equation}
n(r) = n_{\rm cen} ( 1 - r/R_{\rm c})^{-2/5},
\end{equation}
where $T_{\rm cen}$ and $n_{\rm cen}$ are the central temperature and
density of the bubble respectively. The internal structure of the bubble,
and hence the X-ray emission, can be calculated at each instant in such
a semi-analytical model as $T_{\rm cen}$ is a simple function
of the current bubble pressure $p$, age $t$, radius $R_{\rm c}$, and
the coefficient of classical conductivity $C \approx 6 \times 10^{-7} 
\ergps \cm^{-1} \K^{-7/2}$ (Cowie \& McKee 1977):
\begin{equation}
T_{\rm cen} = A \left( \frac{p R_{\rm c}^{2}}{t C} \right)^{2/7}.
\end{equation}
The numerical constant $A = 1.646$ is taken from Weaver \etal (1977).
Again this method assumes that the bubble rapidly comes into 
conductive equilibrium.

These semi-analytical methods are more flexible than the purely analytical
models described in (i) above, and can even be used when the density
distribution is not spherically symmetric, but again do not incorporate
important features such as hydrodynamical instabilities.

\item Computational hydrodynamical models: Although computationally far more
expensive than the two methods described above, numerical hydrodynamical models
are much more flexible. Gravity, radiative cooling, instabilities, time-varying
energy injection, and complex density distributions can all be included.

Their main disadvantages are their computational cost, limited physics
(\eg typically conduction is not included in multidimensional hydrodynamic 
codes), and finite numerical resolution.
\end{enumerate}

We have chosen to investigate the evolution young superbubble blown by
a super star cluster, with time varying mass and energy injection 
using the 2-D  hydrodynamic code {\sc
VH-1}. VH-1 is a third order scheme based on the  Piecewise Parabolic
Method of Colella \& Woodward (1984), with excellent shock handling
and low numerical diffusion characteristics.  See Strickland \&
Stevens (1998) for more discussion regarding VH-1, and X-ray emission
from wind-blown bubbles.

The simulations described in this paper
only investigate a relatively simple model (see
\S~\ref{sec:bubsim}), of a single superbubble evolving into
 a constant density medium. Hence the semi-analytical models described above
can be used to provide an important independent check on the results
of the hydrodynamical simulations, and to
quantify the effect of thermal conduction, which VH-1 does not include
(see \S~\ref{sec:lxfactors}).
This semi-analytical method was used to produce
Figs.~\ref{fig:lh95}b \& d  (bubble radius and shell velocity as a function of
time), given the same model parameters as used in the
hydrodynamic simulation described in \S~\ref{sec:bubsim}.
Future work on more complex hydrodynamical situations, 
for example the interaction and
evolution of multiple superbubbles, can only be done with multidimensional
numerical hydrodynamics.

\begin{figure*}
\vspace{17cm}
\includegraphics{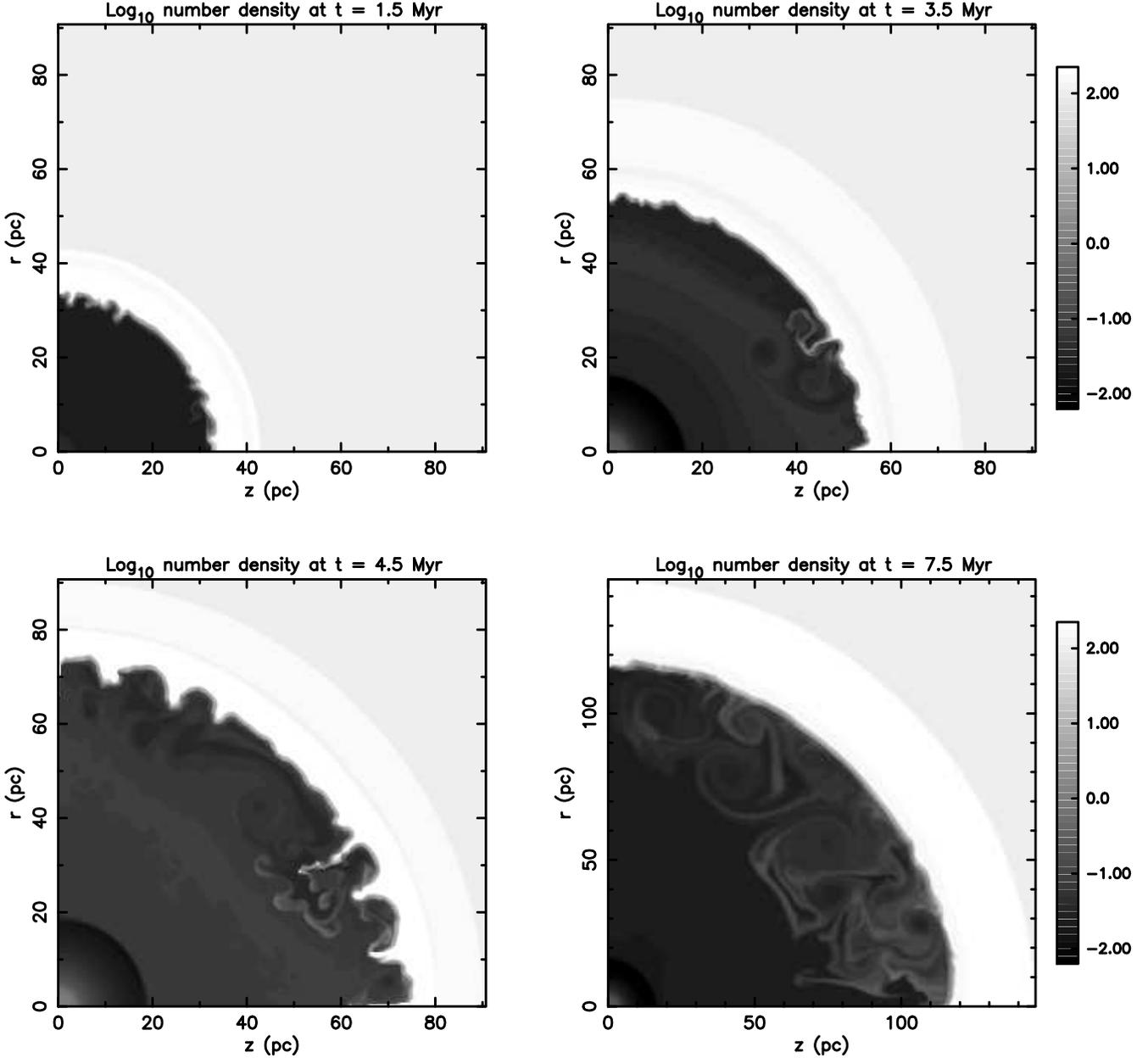}
\caption[Superbubble density structure at four epochs]
 {Superbubble density structure at four epochs representative of the
 structure during the initial 10 Myr, for the
 simulation described in Section~\ref{sec:bubsim}. The bubble structure at 
 $t=1.5\Myr$ is typical of the first 3 Myr, when the mass and energy
 injection is supplied by stellar winds only, and the hot bubble interior
 and cold shell are quite distinct. The additional thermal energy in the
 bubble interior from the first SNe occurring at $t \gtsimm 3 \Myr$ compresses
 and accelerates the superbubble shell outwards between $t \approx 3.5$
 -- $4 \Myr$. The early stages of the
 resulting Rayleigh-Taylor (R-T) instability in the shell can be seen at 
 $t=3.5\Myr$, while at $t=4.5\Myr$ the instability is at its most developed
 stage. After $t \approx 4 \Myr$ the shell is once again R-T stable,
 and the existing R-T ``fingers'' are slowly destroyed and mixed into the
 hot interior by turbulent motions within the bubble. By $t = 7.5 \Myr$
 the shell is smooth and only regions of enhanced density remain 
 within the bubble interior.}
\label{fig:simulation}
\end{figure*}

\subsubsection{Bubbles blown by individual clusters}
The standard picture of starburst driven superbubbles and galactic winds
is that energy and mass injection over a starburst region, typically several
hundred parsecs in size, is efficiently thermalised, creating a hot central
region out of which the hot gas expands and accelerates,
as depicted in Fig.~\ref{fig:bubble_struct}. This is based on
Chevalier \& Clegg's (1985) model for spherically symmetric galactic winds,
since used widely in theoretical (\eg Heckman, Armus \& Miley 1990) 
and hydrodynamical 
(\eg Tomisaka \& Ikeuchi 1988; Tomisaka \& Bregman 1993; Suchkov \etal 1994) 
work. Chevalier \& Clegg assumed the mass and energy injection 
was uniformly distributed over the starburst region. 
Observationally we know a significant fraction of the massive stars
in starburst regions are associated with associations and compact clusters.
Meurer \etal (1995) estimate $\sim20$\% of the 2200 \AA $\,$ light in NGC 5253
comes from the clusters. As a result the mass and energy injection rate per
unit volume will vary strongly within the starburst region.

We can estimate the
injection rates over the entire starburst region 
from \S~\ref{sec:supernovae}. Ignoring stellar winds, then the average 
energy injection rate per unit volume
$Q_{\rm E}$ is simply related to the SN rate $N_{\rm SN}$ and the radius 
of the region considered, by
$Q_{\rm E} = 3 \, N_{\rm SN} / 4 \pi R^{3}$.

For the entire starburst region, 
$Q_{\rm E} \approx 2 \times 10^{-10}$ SN $\pyr \pc^{-3}$, 
where we have taken the
total SN rate to be $N_{\rm SN} = 10^{-3} \pyr$ from \S~\ref{sec:supernovae} 
and used the
effective radius of the starburst region of $100 \pc$ from Meurer \etal (1995).
The true volume of massive star formation
within NGC 5253 is quite a bit larger than the effective radius used above,
as can be seen from Fig.~\ref{fig:xhst}, where it covers a projected
area of $\sim 400\pc$ square, so $Q_{\rm E}$ may be an order of magnitude
lower.

We can use the observed sizes and masses/stellar populations  of the
SSCs from the HST imaging observations of Meurer \etal (1995),
Gorjian (1996)  and Calzetti \etal (1997) and the long-slit
spectroscopic work of Schaerer \etal (1997) (summarised in
Table~\ref{tab:cluster_mass}) to estimate the SN rate in individual
young clusters.

A cluster of total mass of $10^{5} \Msol$, assuming a Salpeter IMF extending 
between $0.1$ -- $100 \Msol$, has a roughly constant SN rate 
$N_{\rm SN} \sim 4 \times 10^{-5} \pyr$ between the ages of
$3.5$ -- $30 \Myr$ and an effective radius 
$R \sim 3 \pc$. Hence $Q_{\rm E} \approx 4 \times 10^{-7}$ SN $\pyr \pc^{-3}$,
several orders of magnitude higher than the average energy injection rate
over the entire starburst region. 

Similarly the mass injection rates into the clusters will be several orders
of magnitude greater than the average mass injection rate into the entire 
starburst region, and the resulting gas pressure will be similarly greater.
As a result individual clusters should blow strong winds into the surrounding
regions of the starburst, even if the entire starburst region is within
a single superbubble cavity.

Depending on the separation between individual massive clusters, and the 
history of star formation in the vicinity, there may or may not be periods
during which each cluster blows its own superbubble before bubbles meet 
and merge.

\subsubsection{Superbubble simulation}
\label{sec:bubsim}

The purpose of this simulation is to obtain an order-of-magnitude
estimate of the X-ray luminosity of a young superbubble as a function of
time. Given the uncertainty in the masses and ages of the clusters in
NGC 5253, and the state of the (multi-phase) ISM, a detailed
comparison of the  observed X-ray emission with hydrodynamical
modeling requires exploring a large parameter space.  As discussed
below in \S~\ref{sec:lxfactors}, 
there are several factors that can alter   the X-ray
luminosities inferred from the HRI observation and those from our
simulations.  We therefore feel it sensible to proceed in a
stage-by-stage basis,  first establishing if a simple model can
provide approximately  the correct X-ray luminosity, and leaving more
a detailed modeling study  of young superbubbles, including additional
physics such as  a non-uniform ISM and mass loading by clouds, to
a later date. 

We simulated the first $10 \Myr$ of the evolution of a  bubble blown
by a cluster of stars of total mass $M_{\star} = 10^{5} \Msol$,
assuming a Salpeter IMF extending between  $0.1$ -- $100 \Msol$. This
mass is typical of the estimated masses for several of the brightest
clusters in NGC 5253  (see Table~\ref{tab:cluster_mass}). We assume
all the stars within the cluster 
form at the same time, and scale the masses and energy
injection rates for quarter  Solar abundance from LH95 down to this
mass. The ISM is assumed to be ionised, uniform and stationary over
the region of interest,  with an ambient number density of $n_{0} = 100
\pcc$ and temperature $T_{0} = 10^{4} \K$. Radiative cooling for the
assumed metallicity of $Z = 0.25 \Zsol$ was incorporated using a
parameterised form of the Raymond-Smith 
emissivities that can be scaled to a chosen metallicity.  A minimum
temperature of $10^{4} \K$ was imposed to prevent over-cooling, and to
represent  heating of the gas by the  strong radiation field from the
massive stars in the starburst region.  Gravitational and magnetic
forces were assumed  to be negligible in comparison to the purely
hydrodynamical forces at work. Mass and energy were injected with each
computational time step within the radius of cluster,  $R_{\star} = 1.3
\times 10^{19} \cm \approx 4.2 \pc$. The computational grid comprised
$225 \times 225$ uniformly sized zones, covering a  physical region
$4.5 \times 10^{20} \cm \approx 146 \pc$ square, and run in
cylindrical coordinates assuming rotational symmetry around  the
x-axis.

In addition to this baseline simulation we describe, we performed two
other simulations. To investigate how our results
depend on numerical
resolution we ran a simulation with the same model
parameters at the basic model, except computational cells half the size
and only covering $2.8 \times 10^{20} \cm \approx 90\pc$ square.
This allows to quantify the effects of numerical
resolution on the derived X-ray luminosities (see \S~\ref{sec:lxfactors}). 
To study how the wind mechanical luminosity
$L_{\rm W}$ affects the X-ray emission we ran a simulation
with a more massive star cluster, $M_{\star} = 10^{6} \Msol$, 
with otherwise the same model parameters as the first
simulation. 
Both of these secondary simulations
evolve off their computational grids at $t \sim 5 \Myr$.

Note that we assume a high ambient number density $n_{0} = 100 \pcc$.
The resulting ISM pressure of $P/k = 10^{6} \K \pcc$ is much higher
than the typical pressure of the undisturbed ISM in the Galaxy, $P/k
\sim 3000 \K \pcc$, but pressure within  starburst regions is known to
be significantly higher than  in the normal ISM (see Heckman \etal 1990). This
value of $n_{0}$ was  chosen as much for computational reasons, as
discussed below, as for scientific reasons. High density gas from the
parent molecular cloud undoubtedly surrounds young clusters, although
its structure is very complex as can be seen in local star forming
regions such as Orion.

We can crudely estimate the average  ISM conditions in the center of
NGC 5253 from the literature: Kobulnicky \& Skillman (1995) report a
peak beamed averaged \hi~column density of $2.6\times10^{21} \pcm2$
towards the core.  If we assume a filling factor of unity and a path length
similar to the beam size of $l = 1 \kpc$, then the volume averaged
$n_{\rm H} \approx \nH / l \sim 1 \pcc$.  Martin (1997) estimates the
electron density in warm photoionised gas at  the centre of NGC 5253
to be $n_{e} = 246^{+49}_{-6} \pcc$, from optical spectroscopy and imaging,
with a filling factor of $\approx 0.01$.  This does provide evidence
of a high thermal pressure  in the warm phase of NGC 5253's ISM.
Similarly high ambient densities and pressure have been  inferred from
observations of the 30 Doradus giant \hii~region in the LMC (see
\S.~\ref{sec:30dor}).

This simulation is computationally challenging to run on a
workstation.  The problem is  following the evolution of the
superbubble for $10 \Myr$, given the characteristic sizes and
time-scales involved.  The computational time step is determined by
the Courant condition,  $dt = C \times dx / {\rm max}(v, c_{\rm s}),$
where $C = 0.5$  is the Courant number, $dx$ is the cell size and
${\rm max}(v,c_{\rm s})$ is the maximum velocity or sound speed on the
grid.  For much of the simulation the sound speed within the energy
injection  region $c_{\rm s} \sim 1500 \kmps$ is the maximum
velocity.  The typical size of the energy injection region is a few
parsecs, given that the half-light radii of the observed clusters are
$\sim 3 \pc$, and this radius must be covered by several ($\gtsimm 6$)
computational cells to give a  reasonable approximation to a spherical
cluster. For the computational grid used above, $dt \approx 210 \yr$,
hence we require $\sim 5 \times 10^{4}$ computational time steps to
cover $10 \Myr$. To keep the runtime to a  sensible level, even on a
fast workstation, requires using as small a number of cells as
possible. As the radius of the contact discontinuity  $R_{\rm c}
\propto n_{0}^{-1/5}$ for fixed $t$ and $L_{\rm w}$, increasing  the
ambient density results in a smaller superbubble and hence less cells
are required in the simulation.

An alternative solution to this problem used by Mac Low \etal (1989;
Mac Low \& Ferrara 1998) is to artificially reduce the maximum
velocity on the grid by increasing the mass injection rate by some
factor  while keeping the energy injection rate constant. This
increases $dt$, and lowers the number of  computational steps
required. It can also be used to roughly approximate the effect of
conduction in increasing the density of the bubble interior.  We chose
not to use this approach, as the density structure of the  bubble
interior with conduction is not known for  time varying energy
injection rates, and it will strongly affect the X-ray luminosity due
to the density squared dependence of $L_{\rm X}.$ Conduction-driven
evaporation off the cold shell is an outside-inwards process, and it
is not immediately apparent if Mac Low \etal's method of additional mass
injection inside-outwards is a reasonable substitute.

For the first 3 Myr the energy injection rate into the  superbubble is
approximately constant at  $L_{\rm w} \approx 10^{38} \ergps$. The
radius of the contact discontinuity $R_{\rm c}$ and the velocity of
the shell  agree well with the thin-shell solutions shown in
Fig.~\ref{fig:lh95}.

The shell of swept-up ISM is well resolved in our simulations,  and is
relatively thick. Densities in the shell are only a few times that of
the ambient ISM, due to the low Mach number $\cal M$ of the forward shock.
This is a consequence of our high chosen value for the ambient 
number density $n_{0}$, as ${\cal M}
\propto n_{o}^{-1/5}$.

Between $t = 3$ and $7 \Myr$ the most massive O stars evolve off the
main sequence and pass through a Wolf-Rayet phase, before going
SN. Note that both clusters G1 and G2 are currently in this WR
phase (Schaerer \etal 1997). 
The first SNe very rapidly increase the wind luminosity to
$L_{\rm w} \approx 10^{39} \ergps$.  The increase in energy injection
is matched by an almost  equal increase in the mass injection rate,
with the result that  the temperature of the bubble interior does not
change significantly.

The sudden increase in thermal pressure of the hot bubble from $t = 3
\Myr$ accelerates and compresses the shell. (Fig.~\ref{fig:lh95}d \&
\ref{fig:simulation}). The acceleration of the shell causes it to
become Rayleigh-Taylor (R-T) unstable between $t \sim 3.5$ and $4
\Myr$, and  R-T fingers of cool shell material penetrate the hot
bubble before being mixed in and destroyed. Note that the growth rate
of the R-T instability is inversely proportional to the square root of
the shell thickness (\cf Tenorio-Tagle \etal 1997), 
so that for lower ambient density $n_{0}$ more shell
material will enter the bubble interior, which may increase the X-ray
emission as discussed below.

After $t = 5 \Myr$ the shell is decelerating once more, and is R-T
stable.  The R-T fingers are mixed into the bubble interior by a
system of  swirling, turbulent motions initiated by the interaction
between the corrugations in the bubble-shell interface and the
slightly faster moving (than the shell) bubble interior 
(Fig.~\ref{fig:simulation}).

\subsubsection{X-ray emission as a function of time}

We calculated the X-ray emission from the bubble in the {\it ROSAT}
 band within the hydro-code itself using a parameterised form of the
 Raymond-Smith $0.1$ -- $2.4 \keV$ X-ray emissivities  that can be
 scaled with metallicity, accurate to  within a few percent.  For each
 cell the product of  $n_{\rm e} \, n_{\rm H} \, \Lambda_{\rm X}(T,Z)$
 was calculated, and multiplied by the volume of each cell using the
 rotational symmetry around the x-axis.  The sum over the entire
 computational grid gives the estimated X-ray luminosity, as shown in
 Fig.~\ref{fig:lh95}c.

Comparing the mechanical energy injection rate with the X-ray emission in
the {\it ROSAT} band (Fig.~\ref{fig:lh95}a \& c), it is clear that
they are very similar in shape,  although there is a slight time lag
between changes in $L_{\rm w}$ and a corresponding alteration in
$L_{\rm X}$.  $L_{\rm X}(t)$ shows the same dramatic increase at  $t
\sim 3 \Myr$ as the energy injection rate does,  peaking at $L_{\rm X}
\approx 9 \times 10^{37} \ergps$,  and settling down to a constant $4
\times 10^{37} \ergps$ after $t = 8 \Myr$. The X-ray luminosity is at
its maximum during the Wolf-Rayet phase between $3$ and $7 \Myr$.  The
superbubble converts a few percent,  $L_{\rm X}(t) \approx 0.05 \times
L_{\rm w}$, of the input mechanical luminosity into soft X-ray
emission.

These X-ray luminosities are remarkably similar to those inferred for
the marginally extended X-ray sources seen by the HRI in NGC 5253
($L_{\rm X} \approx 2 $ -- $ 7 \times 10^{37} \ergps$). This is
probably a coincidence, as there are a variety of factors that can alter
the X-ray luminosity of the superbubble, and the values inferred from
X-ray observations.

\subsubsection{Factors affecting $L_{\rm X}$}
\label{sec:lxfactors}
The X-ray luminosities derived in our simulations depend on the  model
parameters and the physics used. The X-ray luminosities derived from
our HRI data and other worker's observations also suffer from a
major systematic source of uncertainty not reflected by the
statistical errors (see Strickland \& Stevens 1998). 
Although difficult to quantify, it is  worth
discussing these uncertainties and how they might alter both the
simulated and observed X-ray luminosities.  

{\bf Conduction:} Our code does not include the effects of
conduction-driven evaporation off the cold shell into the hot bubble.
Conduction will increase the soft X-ray luminosity of young
superbubbles,  as the density inside the hot shocked wind is increased
by mass from the shell entering and  cooling the bubble interior from
a few $\times 10^{7} \K$ to a few  $\times 10^{6} \K$. Note that the
X-ray emissivity within the {\it ROSAT} band peaks at $\sim 10^{6.8}
\K,$ so part of the luminosity increase will come from increased X-ray
emissivity, although the major effect is due to increasing the number density
of the X-ray emitting gas as  $L_{\rm X} \propto n_{\rm e}^{2}$.

Using the semi-analytical method described in \S~\ref{sec:superbubbles}
we find the {\it ROSAT}-band X-ray emission in a conduction-dominated
superbubble is very similar in form to that found in our hydrodynamical
simulations, although always more luminous. The $0.1$--$2.4 \keV$
luminosity peaks at exactly the same time as in our simulations,
with $L_{\rm X} = 2 \times 10^{38} \ergps$. At later times, for 
$t \geq 7 \Myr$, the X-ray luminosity 
$L_{\rm X} \approx 1.3 \times 10^{38} \ergps$ is approximately constant.  
 
In general this simple conductive model (which does not incorporate 
hydrodynamical instabilities such the R-T instability, which will increase
the X-ray luminosity through the introduction of 
more mass into the interior and an increased conduction evaporation rate)
the X-ray luminosities are typically between 2 -- 4 times that in our
hydrodynamical simulation.

{\bf Numerical resolution and diffusion:} With hydrodynamical models 
it is always important to investigate how the results depend on the numerical 
resolution of the simulation. In models of superbubbles numerical
diffusion can lead to unphysically large amounts of shell material
entering the hot bubble interior, and hence giving unphysically high
luminosities.

We could only follow our higher resolution simulation until $t = 5 \Myr$,
but the qualitative structure and behaviour of the superbubble during
this period were very
similar to the default lower resolution simulation that we have described
in detail. The general features of these simulations are therefore not
strongly affected by the numerical resolution.

The $0.1$--$2.4 \keV$ X-ray luminosities in the higher resolution
simulation were typically only $\sim 60$\% 
of those in the default simulation. This indicates that the structure
of the default bubble is not fully resolved. The simulation is not
so under-resolved as to be unphysical, as the derived X-ray luminosities are
always lower than those derived from the conductive semi-analytical model
mentioned above. 

{\bf Mass-loading:}  Hydrodynamical ablation and destruction of dense
clouds overrun by the bubble will also increase the X-ray luminosity
by increasing the density within the bubble interior, although the
magnitude of this effect requires  investigation. Large amounts of
dense molecular gas left over from the  parent molecular cloud can be
seen in young Galactic star formation regions, \eg Orion (see Wisemann
\& Ho 1996 and references therein). Radio observations of the classic
starburst galaxy M82 suggests the presence of clumpy,
high density clouds (\eg Allen \& Kronberg 1998) within the starburst region.  
Mass-loading is believed to be
important in understanding the X-ray luminosity of SN remnants in the
LMC (\eg Arthur \etal 1993). 

{\bf Cluster mass:} The energy injection rate $L_{\rm w}$ (and hence
$L_{\rm X}$) scales linearly with the  mass of the star cluster. From
Table~\ref{tab:cluster_mass}  it is clear that there is some
uncertainty in the cluster masses, mainly due to different assumptions
concerning cluster ages and optical extinction. For example, Meurer
\etal (1995) assume all the clusters suffer $1.52$ magnitudes of
extinction at $\lambda = 220 \nm$, and are at their peak UV
brightness. Calzetti \etal (1997), in a  multi-wavelength HST study,
find very patchy extinction, reaching $A_{\rm V} = 9$ -- $35$
magnitudes towards the brightest cluster G1. 

Compared to our basic model of a $10^{5} \Msol$ cluster, a pessimistic
view would be that the individual cluster masses may be up an order of
magnitude higher or lower than that used.

Based on the comparison between the $M_{\star} = 10^{5} \Msol$ cluster
and the $10^{6} \Msol$ simulation the X-ray luminosity is almost
directly proportional to $L_{\rm W}$, and hence the cluster
mass, $L_{\rm X} \propto M_{\star}$. Experiments with the semi-analytical
conductive bubble model confirm that the X-ray luminosity is almost
directly proportional to  $L_{\rm W}$ and hence $M_{\star}$.

More X-ray luminous superbubbles can be obtained simply by increasing
the mass of the star cluster, but bear in mind that the X-ray
observations also  place limits on the size of these putative
superbubbles,  and the size depends on the energy injection rate
$R \propto L_{\rm w}^{1/5}$.

{\bf Cluster ages:} Our simulations only cover the first $10 \Myr$,
before the shell evolves off the computational grid, so the X-ray
luminosities for older bubbles  are uncertain. The energy injection
rate remains almost constant until an age of $\sim 30 \Myr$ before
declining as the least massive stars that undergo core collapse die. 
The major effect that will affect $L_{\rm X}$ after $t = 10 \Myr$ will be
changing ambient density, as the bubble expands, and potentially
interacts with other bubbles and cavities in the  ISM.

The ages for clusters G4 \& G5 of $\sim 50 \Myr$  derived by Calzetti
\etal (1997) are problematic for explaining HRI component E by a
superbubble.  Assuming $n_{0}$ and $L_{\rm W}$ remain constant,  the
radius of a $50 \Myr$ old superbubble is 2.6 times that of a  $10 \Myr$ 
old bubble, \ie $\sim 400 \pc$ for the parameters  used in the
hydrodynamic simulation. A bubble this large would subtend $20\arcs$
on the sky at the distance of NGC 5253, much larger than component E!
Lower cluster masses and hence lower energy injection rates cannot be
used to avoid this problem, as $L_{\rm X} \propto L_{\rm W}$, and the
bubble radius is less sensitive to changes in $L_{\rm W}$ or $n_{0}$
than changes in age (Eqn.~\ref{equ:n5253_radius}). 

We speculate that the ages inferred for these clusters have been
over-estimated.  For clusters G1 \& G2 Calzetti \etal use \halpha~and
${\rm H}\beta$  equivalent widths to estimate  the cluster ages,
which agree  well with those derived in  Schaerer \etal's (1997)
spectroscopic study. For the other clusters Calzetti \etal use colours
to estimate the age. Such colour-based age diagnostics do become less sensitive
with increasing age, as shown in their Fig.~8.  Spectroscopic
observations, which have been shown to be of great use in explaining
the complex star formation history  of the SSCs in NGC 1569
(Gonz\'alez Delgado \etal 1997), have yet to be done on all the
clusters in NGC 5253.  This would be of great use in determining the star
formation history of such an important starburst galaxy.

{\bf Ambient Density:} We used a high ambient density of $n_{0} = 100
\pcc$.  From the standard Weaver model,  $L_{\rm X} \propto
n_{0}^{17/35}$, so for an ambient density of $1 \pcc$, the X-ray
luminosity will be approximately an order  of magnitude lower than our
simulation.

Note that for lower $n_{0}$ the shell is more R-T unstable, as the
shell will be thinner and the growth time for the R-T  instability is
proportional to the square root of the shell thickness.  This should
introduce more mass into the bubble interior, which should act to
increase the soft X-ray luminosity to some degree, \eg by enhanced
conductive evaporation into the hot interior from R-T fingers
penetrating into the bubble.

{\bf Metallicity:} SN ejecta and stellar winds from the evolved
massive stars will enrich the superbubble with freshly synthesised
heavy elements.  Conduction and mass-loading will introduce ambient,
\ie less processed, less enriched material, reducing the
metallicity of the X-ray emitting gas. An exact calculation of the
metal abundances is outside the scope of this work, but we assumed a
metal abundance  (relative to Solar) of $Z = 0.25 \Zsol$ for the hot
X-ray emitting gas,  slightly higher than the value of $Z \approx 1/5 \Zsol$
from Walsh \& Roy (1989).

Line emission forms a significant part of the X-ray emissivity of a
hot plasma in the {\it ROSAT} band. To a rough approximation
$\Lambda_{\rm X}(T,Z) \propto Z$, so if the metal abundance (in
particular the Iron abundance for emission within  the {\it ROSAT} band) is
significantly different from $0.25 \Zsol$ the predicted superbubble
X-ray luminosity from Fig.~\ref{fig:lh95}c will need to be scaled
appropriately. 

{\bf Non-ionisation equilibrium (NIE):} The Raymond-Smith emissivities
(1977) used in calculating the X-ray emission and the total
cooling curve used in the simulation assume the plasma is in collisional
ionisation equilibrium. This state is not necessarily achieved throughout
superbubbles and galactic winds.

The soft X-ray emission from a superbubble will be dominated by
initially cool gas mixed into the bubble interior and heated by
conduction, mass loading and instabilities. The ionisation state of
this gas will lag behind its temperature rise, as discussed by Weaver
\etal (1977).  
As a result the ions that in collisional ionisation equilibrium (CIE)
contribute significantly to the X-ray emission of plasmas  with $T
\sim 10^{6} \K$ will in practice be under-ionised. In general then the
true X-ray luminosity will be less than that obtained assuming CIE.

For a suddenly heated plasma, the time required to
reach ionisation  equilibrium, $t_{\rm ieq} \approx 0.03 n_{\rm
e}^{-1} \Myr$ (Masai 1994), which for the gas dominating the 
intrinsic X-ray emission in our simulations 
(using ionisation equilibrium emissivities) 
is typically $10^{-3}$ -- $3 \times 10^{-2} \Myr$, 
significantly shorter than the age of the superbubble. This suggests
NIE effects will not be a major effect for the soft X-ray emission
in this stage of superbubble evolution.

{\bf Transient increases in $L_{\rm X}$ due to SN:} SNRs expanding
within the tenuous bubble will not be X-ray luminous individually due
to the low density,  and it is assumed that their kinetic energy is
thermalised within the bubble.  SN can then be treated as a continuous
form of energy injection, rather than individual events, and this is
the approach taken in the majority of the literature. This assumption
depends on the SN blast wave being subsonic by the time it reaches the
superbubble's shell. Mac Low \& McCray (1988) find that the fractional
radius where a SN blast wave becomes subsonic (assuming the SN
occurred at the centre of the bubble) is  $r/R_{\rm c} \approx
0.47(E_{\rm SN}/U_{\rm bub})^{1/3}$, where $E_{\rm SN}$ is the total
SN energy, $U_{\rm bub}$ is the thermal energy in the
superbubble and $R_{\rm c}$ the radius of the contact discontinuity at the
outer edge of the hot bubble interior. 
Thus, by the time one or two SN have occurred within the
superbubble, all further SN blast waves from the central cluster with
be thermalised within the superbubble. Only SN that occur near the
edge of the superbubble will not be thermalised before they strike the
shell.

Off-center SN blast waves striking the shell while still supersonic,  causing
brief ($\ltsimm 10^{4} \yr$) increases in X-ray luminosity, may be
required to explain some X-ray bright superbubbles around OB
associations in the LMC (Chu \& Mac Low 1990; Wang \& Helfand
1991). Chu \& Mac Low use a simple semi-analytical
model to study the increase in
$L_{\rm X}$ due to this process, but the maximum increase they found
was $\sim  10^{36} \ergps$, which is negligible in terms of explaining
the X-ray emission from NGC 5253.

{\bf Observed X-ray luminosities:} The X-ray luminosities for the five
components given in  Table~\ref{tab:model_5ps} assume that the
spectrum is well represented by a single temperature hot plasma
model. In reality the plasma inside the bubble will be multi-phase
due to density variations within the isobaric bubble interior.

As discussed in Strickland \& Stevens (1998), properties inferred from
simple spectral model fits to multi-temperature X-ray spectra can
deviate significantly, by up to an order of magnitude, from the true
properties.  Although the results of the {\it ROSAT} PSPC spectral
fitting can be  considered as characterising the spectral information,
the X-ray luminosities quoted in Table~\ref{tab:model_5ps} should be
considered more uncertain than the formal error estimates.

\subsection{A comparison with 30 Doradus}
\label{sec:30dor}
The nearest example of a star formation event approaching the
intensity of that creating the clusters in NGC 5253 is 30 Doradus in
the LMC. It is interesting to compare the X-ray properties of 30
Doradus, as inferred from {\it ROSAT} HRI and PSPC observations, with
those of the individual components in NGC 5253.

Optically, the 30 Dor (Tarantula) nebula is $300 \pc$ in diameter
(Chu \& Kennicutt 1994).  At the distance of NGC 5253 it would subtend
$15\arcs$ on the sky. The star formation history within the nebula is
acknowledged to be complex, but the central cluster R136 is probably
young, $3$ -- $5 \Myr$ old, with estimates of  the initial mass of
$\gtsimm 1.7 \times 10^{4} \Msol$ (Malumuth \& Heap 1994), or $\sim 3
\times 10^{4} \Msol$, upper limit $1.5 \times 10^{5} \Msol$ (Brandl
\etal 1996).

The nebula is a source of diffuse  soft thermal ($kT =
0.34^{+0.07}_{-0.04} \keV$) X-ray emission that peaks in holes in the
optical emission, with a total luminosity between $3$ and $6 \times
10^{37} \ergps$ (Norci \& O\"gelman 1995).  A nearby supernova
remnant, N157B, has a X-ray luminosity of  $8 \times 10^{36}
\ergps$. {\it ROSAT} HRI observations reveal two point sources
associated with the central R 136 cluster of luminosity $L_{\rm X}
\sim 10^{36} \ergps$ each, that may be high mass X-ray binary black
hole candidates (Wang 1995). 

On the basis of the size and velocities of individual shells and filaments,
Chu \& Kennicutt (1994) derive ambient number densities in the range 
$n_{0} \approx 5$ -- $70 \pcc$, or $\approx 12 \pcc$ considering 
the entire nebula.

\subsection{Summary of discussion}
\label{sec:summary_disc}
To summarise, the possible sources of X-ray emission that can match 
the luminosities inferred from the HRI observations, and can be 
associated with massive stars are MXRBs, SNRs and superbubbles. We 
would expect all of these sources to be present in NCG 5253 at some
level, given the example of 30 Doradus, but the real question is which
type of source is the {\em dominant} X-ray emitter?

Massive X-ray binaries cannot be the predominant source, primarily for
spectral reasons, as the integrated spectrum from the {\it ROSAT} PSPC
observations is that of a soft thermal plasma, unlike the generally
hard X-ray spectrum associated with MXRBs. This does not rule out
one or two of the observed components, \eg A \& D, or some fraction of the
emission from the extended components, being due to massive binaries,
but summed over all the components the $0.1$ -- $2.4 \keV$ spectrum must
be soft.

Superbubbles and SNRs can both be sources of soft thermal X-ray emission,
although SNRs are generally much fainter than the sources observed in NGC 
5253. For a given X-ray luminosity, SNRs will be much more compact, 
$R_{\rm SNR} \sim 10 \pc$, than superbubbles $R_{\rm c} \sim 50$ --
$150 \pc$. 
To explain the extended HRI components as SNR would require each of the
extended components (B, C \& E, HWHM $\sim 30\pc$ -- $80\pc$)
to be comprised of several smaller SNRs, which begins to run
into problems with the SN rate and the predominantly thermal radio spectrum. 

It is probably inescapable that young star clusters such as
those observed by the HST blow superbubbles into the ISM of the starburst
region -- the main question is can they be X-ray luminous and compact 
enough to explain the observed X-ray components?
We have performed the first hydrodynamical simulations of superbubbles
blown by individual super star clusters with realistic time-varying mass
and energy injection rates. The soft X-ray luminosity $L_{\rm X}(t)$ 
of the bubble was found to be proportional to the mechanical energy injection
rate $L_{\rm W}(t)$, implying that superbubbles should be most X-ray 
luminous between $t = 3$ -- $7 \Myr$, the Wolf-Rayet phase of the parent
cluster. The soft X-ray luminosity for a superbubble blown by a
$M_{\star} = 10^{5} \Msol$ 
cluster during this period and up to $t = 10 \Myr$ agrees very well with
the luminosities inferred for the extended components from the HRI data,
although a variety of factors not considered in our simulations can
influence the absolute magnitude of the X-ray luminosity.

To achieve 
the required X-ray luminosities with superbubbles or SNRs requires them
to be expanding into a region of higher than
average density. 
This may also explain the lack of detected X-ray emission 
from some of the other bright clusters. Considering the count rates of
the observed components, we would not detect sources with 
$L_{\rm X} \ltsimm 5 \times 10^{36} \ergps$, so superbubbles
with lower energy injection rates or
expanding into regions
of lower ambient density could easily have been missed by our
{\it ROSAT} HRI observations. 

The balance of the evidence favours superbubbles as the dominant
source of the X-ray emission from NGC 5253. Only higher resolution 
X-ray spectral imaging by {\it AXAF} can conclusively prove this
to be the case, and identify the origin of the individual components.


\section[Implications for mass loss from dwarf galaxies]
{Implications for mass-loss from NGC 5253 
and other dwarf starburst galaxies}
\label{sec:implications}

What implications do these possible multiple superbubbles
have on starburst-driven mass loss and metal ejection from NGC 5253?
What fraction of the ISM will be blown away, and how efficiently can the
heavy elements synthesised by the massive stars in the starburst escape
into the IGM? The long term fate of the material blown out of galaxies in 
galactic winds is still uncertain, but given the low escape velocities
of dwarf galaxies any hot, metal enriched, 
thermalised SN and stellar wind ejecta that
reaches the halo is probably lost.

The existing X-ray observations of NGC 5253 can not be used to meaningfully
constrain the total gas mass and thermal energy content of the hot ISM.
The X-ray spectrum observed by the {\it ROSAT} PSPC is consistent
with  emission from a soft thermal plasma, but this is not
enough on its own to allow us to infer the properties of what is most
probably, a mix of sources. Superbubbles are theoretically expected to 
have complex multi-temperature X-ray spectra, which as shown in 
Strickland \& Stevens (1998) can give misleading ``derived'' properties
when observed with {\it ROSAT} and fit with simplistic single or 
two-temperature spectral models. A more quantitative investigation of
the properties of the hot gas in NGC 5253 awaits the launch of {\it AXAF},
which can both determine the size of any extended sources, and be used to
characterise the spectral properties of each source.

We shall instead consider qualitatively the implications of the starburst
upon NGC 5253, focusing on analytical and numerical studies of the effect
of starburst-driven superbubbles and winds on dwarf galaxies, \eg De Young
\& Heckman (1994) and Mac Low \& Ferrara (1998).

However, these and other related theoretical work in the literature
(Chevalier \& Clegg 1985; Tomisaka \& Ikeuchi 1988;
Tomisaka \& Bregman 1993), 
which we shall refer to as the ``standard'' model, implicitly assume:
\begin{enumerate}
\item A single superbubble, driven by a single region of uniform mass
and energy deposition, \ie the entire starburst region as a whole drives
a wind into the ISM, \eg Chevalier \& Clegg (1985).
\item A simplified energy and mass injection history, constant with
time for $40$ -- $50 \Myr$, effectively based on one single instantaneous
burst of star formation, and ignoring the weaker energy injection for
the first $3 \Myr $ due to stellar winds alone
\item A single phase, previously undisturbed ISM.
\end{enumerate}

Compare this to the more complex situation in NGC 5253, where we have:
\begin{enumerate}
\item A more complex SF history. 
Super star clusters and
smaller associations  are physically distributed over the starburst 
region, with a range of different ages (although SF within an individual 
cluster may be co-eval), with the possibility that the region of
active SF has moved from the SW to the NW (Gorjian 1996). Each cluster
is probably the source of a strong wind of thermalised SN and stellar
wind ejecta, as shown in \S~\ref{sec:bubsim}. 
Complex SF histories, with multiple
clusters of different ages seem to reasonably common, as this situation
is inferred to be the case in other famous starburst galaxies such as
M82 (Satyapal \etal 1997) and NGC 1569 (Vallenari \& Bomans 1996; 
Gonz\'alez Delgado \etal 1997).
\item A multiphase ISM, with molecular clouds (Turner \etal 1997), 
a very prominent dust lane,
regions of dense warm gas (Martin \& Kennicutt 1995; Martin 1997)
and holes and bubbles of hot coronal gas (this work).
\end{enumerate}

We shall briefly consider the qualitative effect of this more complex
situation on the mass loss and metal ejection efficiency, after considering
the predictions of the standard single superbubble model.

\subsection{Mass loss from galaxies in the standard superbubble model}
The coupling between the energy supplied by the SNe and stellar winds from the
starburst and the ISM is crucial in determining the amount of gas and metal
lost from dwarf galaxies. The energy supplied by the multiple supernovae
is sufficient to eject a large mass of gas from the shallow gravitational
potential well of a dwarf galaxy, but this energy must be transferred from
the thermal energy of the hot bubble interior to kinetic energy in
bulk motions of the ambient ISM.

As a simple estimate of the power of a starburst, a starburst can
accelerate a mass of gas $M_{\rm acc} \sim 10^{4} f \, E_{51}/v^{2}_{100} 
\Msol$ to a velocity of $v_{100}$, 
where $v_{100}$ is the velocity in units of $100 \kmps$,
$E_{51}$ is the total mechanical energy injected by SNe and stellar winds
in units of $10^{51} \erg$ and $f$ measures the coupling between the
superbubble and the ISM that is swept-up. For the Weaver model, with
constant energy injection in a uniform ISM, only about 20\% of the total
energy injected ends as the kinetic energy of the swept-up ISM, 45\%
is stored as thermal energy in the hot shocked wind material that makes up
the bubble interior and which drives the bubble's expansion, and most of the
rest is radiated from the shell. In practice $f = 0.2$ is an upper limit.
A single cluster of $M_{\star} = 10^{5} \Msol$,
typical of the brighter clusters in NGC 5253, will inject 
$\sim 10^{54} \erg$ of mechanical energy into the ISM over the
lifetime of the lowest mass star to go SN, \ie $\sim 30 \Myr$. If we adopt
the total initial mass for the entire starburst from 
\S~\ref{sec:supernovae}
of $\sim 5 \times 10^{6} \Msol$ then $E_{51} \sim 5 \times 10^{4}$.
NGC 5253's total mass is  $\ltsimm 10^{9} \Msol$ ($3 \times 10^{8} \Msol$
of which is gas, Kobulnicky \& Skillman 1995), so the escape velocity
is $\sim 100 \kmps$, using Eqn.~28. of Mac Low \& Ferrara (1998).  
Hence based on the very simple energetics argument above we might expect 
{\em at maximum} $\sim 10^{8} \Msol$ of the ISM to be 
ejected from NGC 5253 over the
entire history of the current starburst.

Once any superbubble breaks out of the disk into the halo, the coupling
of the starburst to the remaining ISM is reduced as the bubble escapes 
the disk and expands preferentially into the much lower 
density and pressure environment 
of any galactic halo medium and the IGM, venting the thermal energy of 
the bubble interior. The hot gas previously forming the superbubble
interior accelerates outwards as a galactic wind, carrying with it
the fragmented remains of part of the dense superbubble shell 
(between 4 -- 9\%
of the shell mass in Mac Low \etal's [1989] simulations of 
superbubbles blown by normal OB associations).
The amount of energy transferred to the ISM then
drops once any effective chimney to the halo is created, and hence 
the amount of gas ejected with the hot, unbound, wind material is also
reduced.

How effectively the starburst couples with the ISM depends on the structure
and distribution of the ISM in the galaxy, as well as the mechanical power
of the starburst. De Young \& Heckman (1994) considered analytically
the effect of a superbubble of constant power, lasting for $40 \Myr$,
expanding in an uniform 
density elliptical ISM for galaxies of different mass, ISM density and
ellipticity.

Although their models did not include massive dark matter haloes
around their galaxies, their results illuminate much of the basic physics,
and are worth reiterating here.
They found that thin disks, dense disks, and low total energy injection
all encourage blow-out along the minor axis. Conversely, thick disks,
tenuous disks and high energy injection all increase the probability of
a catastrophic blow-away of the ISM. Normal massive disk galaxies are
very resistant to blow-away, while low mass dwarf galaxies 
($\sim10^{7} \Msol$) are  susceptible to having their entire ISMs blown
away. Typical dwarf galaxies of $M \sim 10^{9} \Msol$ show a wider range of
possible behaviour, dependent on the distribution of the ISM.

Mac Low \& Ferrara (1998) repeated De Young \& Heckman's study, this
time incorporating the effects of massive dark matter haloes, in an attempt
to obtain quantitative estimates for the total mass and metal ejection 
efficiencies for a range of dwarf galaxy gas masses and mechanical
luminosities, using a 2D hydrodynamic code.
Again the energy injection rate
was assumed to be constant in time, lasting for $50 \Myr$.
The maximum energy injection rate they considered
was $L_{\rm W} = 10^{39} \ergps$, equivalent to only a single $M_{\star} = 
10^{5} \Msol$ cluster, which is unfortunately an order of magnitude
or so lower than the total energy injection rate in a starbursting dwarf
like NGC 5253.
They found that the total mass ejected at greater than escape velocity
is $\sim 10^{5} \Msol$ for a galaxy with a gas mass of 
$M_{\rm gas} = 10^{8} \Msol$, although almost all the metal enriched gas
escapes. 
Although they did not perform a higher mechanical luminosity simulation
appropriate for NCG 5253 and similar starburst galaxies,
their results strongly suggest only a small fraction of the total
ISM mass is lost in the wind but
the majority of the newly synthesised metals are lost.
 
Several authors have reported simulations of the more powerful 
starburst in M82 (Tomisaka \& Ikeuchi 1988; Tomisaka \& Bregman 1993; 
Suchkov \etal 1994, 1997), although they do not give quantitative values for
the mass ejected.

\subsection{Mass loss with realistic star formation histories}
Suchkov \etal (1994) do use a more realistic SF history than previously
considered, using the LH95 models to give the time-dependent 
mass and energy injection
rates due to a constant star formation rate of $2 \Msol \pyr$, and they
stress the importance of the starburst history in altering 
the dynamics of a galactic wind. They found that the weak
initial mass and energy injection, due to stellar winds before any SNe, creates
 a cavity in the disk without strongly affecting the disk ISM.
This allowed the hot gas from the much more powerful SNe dominated phase of the
starburst to escape virtually freely into the halo.

We conclude that a SF history such as NGC 5253's, with the
energy injection spread out more over time due to the range in
cluster ages, will reduce the total mass loss from the galaxy compared
to the standard single superbubble model.
Hence realistic star formation histories (and more realistic, multiphase
ISM distributions) will reduce the efficiency with which starbursts can
strip a galaxy of its ISM compared to the single superbubble model.
Although total mass loss rates may be reduced, easier venting of the 
hot metal-enriched gas from the starburst into the IGM will 
act to increase the metal ejection efficiency.

The reason weaker mass and energy injection cause less disruption to the
ISM is that the total energy per unit volume (or per unit mass) they impart to
the ISM by the time the bubbles blow out is less, even though it may take longer for a weaker superbubble to blow out. Post blow-out, much of the 
energy injected by the starburst, 
including any from later more powerful star formation,
is simply vented directly into the halo without interacting with the remaining
ISM. 

This is due to the size of a superbubble being a stronger power of
time than of wind luminosity $L_{\rm W}$, so much weaker bubbles
need only be slightly older to occupy the same volume as a powerful
superbubble. The total energy supplied to the ISM is roughly 
proportional to $L_{\rm W} t$. In the Weaver model, where the ambient
density and the wind luminosity are constant, then the volume of
the bubble $V \propto L_{\rm W}^{3/5} \, t^{9/5}$ and the kinetic
energy of the swept-up ISM is $E_{\rm ISM} \approx 0.2 L_{\rm W} t$.
For time-varying energy injection and non-uniform ambient density the
standard scaling relationships do not strictly apply, 
and the kinetic energy of the shell is not a constant fraction of the 
total energy injected, but as they are approximately correct
we shall use them to illustrate the point. The energy per unit volume
supplied to the ISM is then $E_{\rm ISM}/V \propto L_{\rm W}^{-2/5} \,
t^{-4/5}$.

Weak superbubbles supply much less energy to the ISM than more powerful
ones do, even though both blow-out after sweeping up the same mass of gas. 
As a result weaker bubbles accelerate the ISM to lower velocities before
blowing out, and with lower ram pressure are less efficient at dragging
dense gas out of the disk into the halo.   

For example consider two cases where the mechanical luminosities
differ by a factor of ten.
If we assume blow-out occurs when any superbubble reaches some critical 
volume determined by the structure of the ISM only, then we can
evaluate both the time taken to blow-out and the energy per unit mass
supplied to the ISM in terms of the wind luminosity. To occupy
the same volume, $t_{1}/t_{2} = (L_{1}/L_{2})^{-1/3}$, a bubble
an order of magnitude weaker would only take approximately twice the time
to blow out, but only supply a fifth as much energy into accelerating 
and disrupting the ISM.

As a result, powerful starbursts that would have sufficient energy to
unbind a significant portion of the ISM of a dwarf galaxy in the absence
of any other star formation, may have little effect on the ISM if preceding
weaker star formation and the resulting superbubbles open up channels to
the halo that efficiently vent the hot gas from the main SF.

Even if we assume no preceding weaker SF within the starburst itself, 
natural levels of massive star 
formation and type Ia SN in galaxies 
can increase the porosity of the ISM and lead to 
significant filling factors of hot, low density gas in the ISM that can
act as vents for starburst-driven superbubbles (\eg see Oey \& Clarke 1997;
Rosen \& Bregman 1995; Rosen, Bregman \& Kelson 1996).

The observational properties of the starburst will also differ from the 
standard single superbubble case, as the structure of the ISM the 
superbubbles interact with will be determined by the history 
of mass and energy injection.
The \halpha~emission, due to both photoionisation and shock heating,
depends on the local SF history, as the ionising flux from young clusters
is a strong function of time (decreasing rapidly after $\sim 5 \Myr$), 
while bubbles both cause shock heating and
allow ionising photons to propagate further out from bright UV sources
(see Martin 1997 for observational evidence and a discussion of this).
The X-ray luminosity of the starburst is also proportional to a weak function
of the ambient density (see Eqn.~\ref{equ:n5253_lx}). Hence $L_{\rm X}$ 
will decrease after blow-out as the galactic
wind encounters only low density halo gas, and as the density of the wind
material decreases as it expands into the galactic halo. The main remaining
regions of X-ray emission will be in the interfaces between hot tenuous 
wind material and any dense gas encountered, \ie near the chimney walls,
shell fragments and clouds the wind shocks and ablates.

\subsection{Summary of implications}
\label{sec:summary_imp}
The total mechanical energy released by the starburst in NGC 5253 is enough
to unbind a significant fraction of its ISM, but the total mass loss
depends crucially on how well this energy can be transferred from the
hot gas to the ISM, before any blow-out and venting of the hot shocked gas
into the halo as a galactic wind.

The SF history in NGC 5253, with multiple clusters of different ages each
blowing winds and bubbles into the ISM, is more complex than the 
single superbubble model with a single instantaneous burst of SF. We
speculate that the total mass ejected from the ISM in a situation with
a SF history similar to NGC 5253's will be reduced compared to the
standard single superbubble model, as the more gradual energy injection
allows the hot gas created by the starburst to blow-out without transferring
as much energy to the ISM. Later thermalised SNe and stellar wind 
ejecta is then vented easily into the halo without interacting significantly
with the remaining ISM. This should also act to increase the fraction of the 
newly synthesised heavy elements that are lost, even though the total mass
lost is reduced. It is unlikely that NGC 5253 will lose a significant fraction
of its ISM to starburst-driven winds, despite this being energetically 
feasible.

\section{Conclusions}
\label{sec:n5253_conc}
We obtained a long {\it ROSAT} HRI observation of the dwarf starburst
galaxy NGC 5253, with the aim of resolving the superbubble believed
to be present on the basis of the {\it ROSAT} PSPC observation of
Martin \& Kennicutt (1995).

\begin{enumerate}
\item Instead of a single superbubble,
HRI observations show several, at least five, separate
sources of soft, 0.1-2.4 $\keV$, X-ray emission, associated with the
massive young clusters of stars at the centre of NGC 5253. 
Based on the PSPC spectrum
the luminosities of the five components range between  $L_{\rm X}
\approx 2$ -- $7 \times 10^{37} \ergps$. 

\item Three of the components are
statistically extended beyond the HRI PSF at 90\% confidence,  the
largest having a FWHM of $8^{+10}_{-4}$ arcsec, equivalent to
$160^{+200}_{-80}\pc$.  

\item The possible sources of X-ray emission that can match 
the luminosities inferred from the HRI observations, and are expected to be 
associated with massive stars are MXRBs, SNRs and superbubbles. We 
would expect all of these sources to be present in NCG 5253 at some
level, given the example of 30 Doradus, but which is the
dominant source of soft X-ray emission?

\item Massive X-ray binaries cannot be the predominant source, primarily for
spectral reasons as the integrated spectrum from the {\it ROSAT} PSPC
observations is that of a soft thermal plasma, unlike the generally
hard X-ray spectrum associated with MXRBs. This does not rule out
one or two of the observed components, or some fraction of the
emission from the extended components, being due to massive binaries,
but summed over all the components the $0.1$ -- $2.4 \keV$ spectrum must
be soft.

\item Explaining the extended HRI components as middle-aged X-ray 
luminous SNRs requires each of the
extended components (B, C \& E, HWHM $\sim 30\pc$ -- $80\pc$)
to be comprised of several smaller SNRs, as for a given X-ray luminosity, 
SNRs will be much more compact, $R_{\rm SNR} \sim 10 \pc$, than 
superbubbles $R_{\rm c} \sim 50$ -- $150 \pc$. This model of several SNRs
per X-ray component
begins to run into problems with the SN rate and the upper limit on the total
number of SNRs being $\ltsimm 10$ from the thermal radio spectrum. 

\item We show that super star clusters and other massive clusters of
stars, such as those observed by the HST in the centre of NGC 5253,
individually  will be the source of strong winds blowing
into the ISM of the starburst region.
These may be the source of {\em multiple} young superbubbles
within the central regions of a young starburst galaxy such as NGC 5253.

\item We have performed the first hydrodynamical simulations of superbubbles
blown by individual super star clusters with realistic time-varying mass
and energy injection rates, in order to investigate whether they can
be X-ray luminous enough to explain the extended soft X-ray components
seen in NGC 5253. The soft X-ray luminosity $L_{\rm X}(t)$ 
of the superbubble was found to be proportional to the mechanical energy injection
rate $L_{\rm W}(t)$, implying that superbubbles should be most X-ray 
luminous between $t = 3$ -- $7 \Myr$, the Wolf-Rayet phase of the parent
cluster. This may explain why Wolf-Rayet galaxies are observed to be
X-ray overluminous compared to normal spiral and more mature starburst
galaxies (see Stevens \& Strickland 1998a; 1998b).
The predicted soft X-ray luminosity for a superbubble blown by a
$M_{\star} = 10^{5} \Msol$ 
cluster during this period and up to $t = 10 \Myr$ agrees very well with
the luminosities inferred for the extended components from the HRI data,
although a variety of factors we discuss that are 
not considered in our simulations can
influence the absolute magnitude of the X-ray luminosity.

\item Achieving 
the required X-ray luminosities with superbubbles or SNRs requires them
to be expanding into a region of higher than
average density. 
This may also explain the lack of detected X-ray emission 
from some of the other bright clusters. Given the count rates for
the observed components, we would not detect sources with 
$L_{\rm X} \ltsimm 5 \times 10^{36} \ergps$, so superbubbles
with lower energy injection rates or
expanding into regions
of lower ambient density could easily have been missed by our
{\it ROSAT} HRI observations. 

\item In conclusion, the evidence favours several of the observed X-ray sources
being young superbubbles, and that superbubbles are the dominant
source of the X-ray emission from NGC 5253. Only higher resolution 
X-ray spectral imaging by {\it AXAF} can conclusively prove this
to be the case, and identify the origin of the individual components.
If confirmed, this is the first detection of multiple superbubbles
in a starburst region.


\item We emphasise the importance of the SF history and distribution
on the amount of gas
and metals ejected from starbursting dwarf galaxies, in addition to the
effect of the ISM distribution that previous work had highlighted 
(\eg De Young \& Heckman 1994; Heckman \etal 1995). 
 
\item We
speculate that the total mass ejected from the ISM of a dwarf
galaxy in a situation with
a SF history similar to NGC 5253's will be reduced compared to the
standard single superbubble model, as the more gradual energy injection
allows the hot gas created by the starburst to blow-out without transferring
as much energy to the ISM. Later thermalised SNe and stellar wind 
ejecta is then vented easily into the halo without interacting significantly
with the remaining ISM. This should also increase the fraction of the 
newly synthesised heavy elements that are lost, even though the total mass
loss is reduced. It is unlikely that NGC 5253 will eject a significant fraction
of its ISM to starburst-driven winds, despite this being energetically 
feasible.
\end{enumerate}

\subsection*{Acknowledgements}

We would like to thank Crystal Martin and Varoujan Gorjian 
for supplying us with optical images of NGC 5253. 
We would also like to thank the following people
for useful discussions and constructive criticism
through various stages of this work: Crystal Martin,
Sally Oey, the members of the Birmingham ``Galaxies and Clusters'' group
and the referee Mordecai-Mark Mac Low.
DKS and IRS also acknowledge the funding support of PPARC and the School of
Physics \& Astronomy. All calculations were carried out on the local 
{\sc Starlink} computing node.

This research has made use of the Leitherer \& Heckman (1995)
starburst evolutionary synthesis models 
(http://www.stsci.edu/ftp/science/starburst/), the
NASA/IPAC Extragalactic Database
(NED, operated by the Jet Propulsion Laboratory, California
Institute of Technology, under contract with the National Aeronautics and
Space Administration), the SIMBAD database (operated by the CDS,
Strasbourg) and the HST Guide Star Catalogue (The Guide Star Catalog 
was produced at the Space Telescope Science Institute under U.S.
Government grant. These data are based on photographic data obtained 
using the Oschin Schmidt Telescope on Palomar Mountain and the 
UK Schmidt Telescope.).

\label{lastpage}
\end{document}